\documentclass[pdftex,12pt,a4paper]{scrartcl}

\usepackage[utf8]{inputenc}

\usepackage[a4paper,left=2cm,right=2cm,bottom=2cm,top=2cm]{geometry}
\usepackage{abstract} 
\usepackage{flafter}
\usepackage{graphicx}

\usepackage{siunitx} 

\usepackage{enumitem}

\usepackage[
	backend=biber,
	url=false,
	doi=true,
	maxcitenames=2,
	style = authoryear,
  hyperref=true,
  uniquelist=false,
	maxbibnames = 99]{biblatex}

\DeclareCiteCommand{\cite}
  {\usebibmacro{prenote}}
  {\usebibmacro{citeindex}%
   \printtext[bibhyperref]{\usebibmacro{cite}}}
  {\multicitedelim}
  {\usebibmacro{postnote}}

\DeclareCiteCommand*{\cite}
  {\usebibmacro{prenote}}
  {\usebibmacro{citeindex}%
   \printtext[bibhyperref]{\usebibmacro{citeyear}}}
  {\multicitedelim}
  {\usebibmacro{postnote}}

\DeclareCiteCommand{\parencite}[\mkbibparens]
  {\usebibmacro{prenote}}
  {\usebibmacro{citeindex}%
    \printtext[bibhyperref]{\usebibmacro{cite}}}
  {\multicitedelim}
  {\usebibmacro{postnote}}

\DeclareCiteCommand*{\parencite}[\mkbibparens]
  {\usebibmacro{prenote}}
  {\usebibmacro{citeindex}%
    \printtext[bibhyperref]{\usebibmacro{citeyear}}}
  {\multicitedelim}
  {\usebibmacro{postnote}}

\DeclareCiteCommand{\footcite}[\mkbibfootnote]
  {\usebibmacro{prenote}}
  {\usebibmacro{citeindex}%
  \printtext[bibhyperref]{ \usebibmacro{cite}}}
  {\multicitedelim}
  {\usebibmacro{postnote}}

\DeclareCiteCommand{\footcitetext}[\mkbibfootnotetext]
  {\usebibmacro{prenote}}
  {\usebibmacro{citeindex}%
   \printtext[bibhyperref]{\usebibmacro{cite}}}
  {\multicitedelim}
  {\usebibmacro{postnote}}

  \DeclareCiteCommand{\textcite}
  {\boolfalse{cbx:parens}}
   {\usebibmacro{citeindex}%
	 \printtext[bibhyperref]{\printnames{labelname}%
	   \printtext{ (\printfield{year}\printtext{)}}}}
  {\ifbool{cbx:parens}
   {\bibcloseparen\global\boolfalse{cbx:parens}}
   {}%
  \multicitedelim}
 {\usebibmacro{textcite:postnote}}

\usepackage[pdftex, colorlinks, allcolors = Blue]{hyperref}

\hypersetup{
  pdftitle    = {Distributional Modeling and Forecasting of Natural Gas Prices},
  pdfsubject  = {We examine the problem of modeling and forecasting European Day-Ahead and Month-Ahead natural gas prices. For this, we propose two distinct probabilistic models that can be utilized in risk- and portfolio management. We use daily pricing data ranging from 2011 to 2020. Extensive descriptive data analysis shows that both time series feature heavy tails, conditional heteroscedasticity, and show asymmetric behavior in their differences.
  We propose state-space time series models under skewed, heavy-tailed distribution to capture all stylized facts in the data.
  They include the impact of autocorrelation, seasonality, risk premia, temperature, storage levels, the price of European Emission Allowances, and related fuel prices of oil, coal, and electricity.
  We provide a rigorous model diagnostic and interpret all model components in detail.
  Additionally, we conduct a probabilistic forecasting study with significance tests and compare the predictive performance against literature benchmarks.
  The proposed Day-Ahead (Month-Ahead) model leads to a $13\%$ ($9$\%) reduction in out-of-sample CRPS compared to the best performing benchmark model, mainly due to adequate modeling of the volatility and heavy tails.},
  pdfauthor   = {Jonathan Berrisch, Florian Ziel},
  pdfkeywords = {probabilistic forecasting, gas prices, natural gas market, state-space models, risk premium, volatility, heavy tailed distribution
  }
}

\usepackage[english]{babel}
\usepackage[utf8]{inputenc}
\usepackage[T1]{fontenc}
\usepackage{ae,aecompl}
\usepackage{csquotes}

\usepackage{eurosym}

\usepackage{amsmath,amssymb}
\usepackage{breqn}
\usepackage{bbold}

\usepackage{url}

\usepackage{booktabs}
\usepackage{tabulary}
\usepackage{tabu}
\usepackage[dvipsnames]{xcolor,colortbl}
\newcolumntype{Y}{>{\centering\arraybackslash}X}
\usepackage{array}
\newcolumntype{P}[1]{>{\centering\arraybackslash}p{#1}}
\newcolumntype{M}[1]{>{\centering\arraybackslash}m{#1}}

\definecolor{s}{rgb}{.8,.85,.8}
\definecolor{sss}{rgb}{.7,.925,.7}
\definecolor{ssss}{rgb}{.6,1,.6}

\definecolor{DA}{HTML}{85B464}
\definecolor{MA}{HTML}{5391AE}

\usepackage[nolist]{acronym}

\usepackage{authblk}

\usepackage[nameinlink]{cleveref}

\definecolor{cyan}{rgb}{0,0.7,.2}
\definecolor{dcyan}{rgb}{0,0.5,.5}

 \definecolor{pink}{HTML}{ff00d4}
 \definecolor{dpink}{HTML}{99007f}

 \newcommand{\DA}[1]{\color{DA} #1 \color{black}}
 \newcommand{\MA}[1]{\color{MA} #1 \color{black}}

\title{Distributional Modeling and Forecasting of Natural Gas Prices}

\author[1]{Jonathan Berrisch}
\author[2]{Florian Ziel}

\affil[1,2]{House of Energy Markets and Finance \authorcr 
University of Duisburg-Essen \authorcr
Universitätsstr. 2,  45141 Essen, \authorcr Germany}

\affil[1]{Jonathan.Berrisch@uni-due.de}

\affil[2]{Florian.Ziel@uni-due.de}

\date{\today}

\begin{document}

\maketitle

\begin{abstract}

  We examine the problem of modeling and forecasting European Day-Ahead and Month-Ahead natural gas prices. For this, we propose two distinct probabilistic models that can be utilized in risk- and portfolio management. We use daily pricing data ranging from 2011 to 2020. Extensive descriptive data analysis shows that both time series feature heavy tails, conditional heteroscedasticity, and show asymmetric behavior in their differences.
  We propose state-space time series models under skewed, heavy-tailed distributions to capture all stylized facts of the data.
  They include the impact of autocorrelation, seasonality, risk premia, temperature, storage levels, the price of European Emission Allowances, and related fuel prices of oil, coal, and electricity.
  We provide rigorous model diagnostics and interpret all model components in detail.
  Additionally, we conduct a probabilistic forecasting study with significance tests and
  compare the predictive performance against literature benchmarks.
  The proposed Day-Ahead (Month-Ahead) model leads to a $13\%$ ($9$\%) reduction in out-of-sample continuous ranked probability score (CRPS) compared to the best performing benchmark model, mainly due to adequate modeling of the volatility and heavy tails.

  \newpage

\end{abstract}

\begin{acronym}[dummyyyy]
  \acro{ACF}{Autocorrelation Function}
  \acro{ADF}{Augmented Dickey-Fuller}
  \acro{ANN}{Artificial Neural Network}
  \acro{AR}{Autoregressive}
  \acro{ARI}{Autoregressive Integrated}
  \acro{ARIMA}{Autoregressive Integrated Moving Average}
  \acro{ARMA}{Autoregressive Moving Average}
  \acro{BAO}{Balancing Area Operator}
  \acro{CAPM}{Capital Asset Pricing Model}
  \acro{CRPS}{Continuous Ranked Probability Score}
  \acro{CSSED}{Cumulative Sum of Squared Predictive Error}
  \acro{DM}{Diebold-Mariano}
  \acro{ETS}{Error-Trend-Seasonal}
  \acro{EUA}{European Emission Allowances}
  \acro{GARCH}{Generalized Autoregressive Conditional Heteroscedasticity}
  \acro{H-Gas}{High-Calorific Natural Gas}
  \acro{IGARCH}{Integrated GARCH}
  \acro{L-Gas}{Low-Calorific Natural Gas}
  \acro{LEBA}{London Energy Brokers' Association}
  \acro{LNG}{Liquified Natural Gas}
  \acro{LSTM}{Long Short-Term Memory}
  \acro{MA}{Moving Average}
  \acro{MAE}{Mean Absolute Error}
  \acro{OLS}{Ordinary least squares}
  \acro{PACF}{partial autocorrelation function}
  \acro{PP}{Philipps-Perron}
  \acro{RMSE}{Root Mean Squared Error}
  \acro{RWEST}{RWE Supply and Trading GmbH}
  \acro{TGARCH}{Threshold GARCH}
  \acro{VAR}{Vector Autoregressive}

  \acro{dummyyyy}{dummyyy}
\end{acronym}


\section{Introduction}\label{introduction}

Natural gas is one of the most important energy sources worldwide. In Europe,
its relevance increased in recent years since it is one of the cleanest fuels, and thus it can be used as a transition fuel during the transition towards a renewable-based energy system. Increasing prices for carbon emissions accelerate this development \autocite{safari2019natural}.

Today's natural gas markets are liberalized to a great extend. In consequence, natural gas became a financial commodity and short-term planning became increasingly important. That is, accurate price forecasts are attractive for various market participants like natural gas field operators, energy companies, and storage operators. These forecasts are utilized for commodity trading, portfolio and risk management, energy sustainability planning, and predictive maintenance.



The literature on natural gas price forecasting grew rapidly in the last decade. Recent studies concerning price forecasting focused on comparing the performance of different methods rather than exploring fundamental relationships between input variables. We want to highlight the major contributions chronologically.

\textcite{nguyen2010short} combined wavelet transformations with various models such as linear regression, \acp{ANN}, and \ac{GARCH} to forecast natural gas prices.
The considered linear models outperformed the nonlinear counterparts like the multilayer perceptron models.
\textcite{salehnia2013forecasting} focused on variable selection and proposed
using the gamma test for this. Afterward, they used several forecasting models, including local linear regression and \acp{ANN}, to predict natural gas prices at different time scales. Their results favor the \ac{ANN} that did not use any external inputs.
The importance of prior variable selection was later revisited by \textcite{vceperic2017short}. They showed that prior variable selection is crucial for obtaining accurate forecasts using \acp{ANN} and support vector regressions. Otherwise, classical time series models outperform these machine learning methods.
\textcite{geng2017relationship} explored the relationship between crude oil and natural gas prices using linear and nonlinear Granger causality. They find that the oil price linearly causes the natural gas price. \textcite{su2019data_a} used linear regression, support vector machines, and boosting to predict daily natural gas spot prices. They followed \textcite{vceperic2017short} by using heating oil prices, storage capacities, and natural gas imports as explanatory variables. The LSboost algorithm, which minimizes the squared error, outperformed the linear regression and the support vector machine. An \ac{ANN} outperformed gradient boosting in a previous study. However, that study focused on forecasting monthly natural gas prices, and an even broader set of explanatory variables was used \autocite{su2019data_b}.
\textcite{herrera2019long} compared a random forest model, an \ac{ANN} and a hybrid model which uses five more traditional econometric methods for forecasting monthly natural gas prices. The authors did not use any external regressors. The results were clearly in favor of the random forest model. \textcite{siddiqui2019predicting} forecasted daily natural gas spot prices using an \ac{ARIMA} model and an autoregressive \ac{ANN}. The \ac{ANN} outperformed the \ac{ARIMA} model, and the model developed in \textcite{salehnia2013forecasting}. \textcite{livieris2020advanced} used a combination of advanced deep learning methods. In particular, they combined convolutional layers with \ac{LSTM} layers to predict the Day-Ahead natural gas prices. They also did not use any external regressors. \textcite{jianliang2020daily} reviewed three common models, i.e., support vector regression, \ac{LSTM}, and the improved pattern sequence similarity search. Afterward, they developed a weighted model that combines the three methods mentioned above. The results showed that the weighted model surpassed the other considered model.

The determinants of the natural gas price were studied by \textcite{wang2019financialization} using dynamic model averaging. Their results show an evident decline in the impact of oil from 2001 to 2018 while fundamental factors like demand, supply, and the weather became increasingly important. The results also revealed an increasing impact of financial markets. This effect was captured by the impact of variables related to financial speculation.
Further, \textcite{tang2019natural} compared the predictive power of google search data and news sentiment to predict the natural gas spot price. Their results showed that using google search data can improve forecasts that otherwise solely rely on past prices.
Recently, \textcite{wang2020forecasting} build probabilistic forecasting models for several commodities, including natural gas. Their study checked whether technical indicators have predictive power. The forecasting accuracy was also compared to models using macroeconomic variables. The results showed that models using a combination of technical indicators outperform models only using economic predictors.


There is also relevant literature on forecasting natural gas demand. \textcite{chen2020day} developed a hybrid model to forecast hourly natural gas demand at 96 distribution nodes across Germany. They combined \ac{AR} models with convolutional \acp{ANN} to reduce overfitting on the one hand but allow nonlinear effects on the other hand.  \textcite{karabiber2020forecasting} also considered \acp{ANN} alongside \ac{ARIMA} models as well as the TBATS model \autocite{de2011forecasting}. They considered many exogenous regressors such as Fourier terms, temperature,  the day type (workdays, weekends, holidays), wind speed, solar radiation, electricity prices, gas prices, and biogas production. The Fourier terms, the temperature, and the lagged consumption were included in all models. The best-performing method combined an \ac{ARIMA} model with an \ac{ANN}. This model surpassed the aforementioned TBATS model.


Finally, there is research on forecasting the volatility of natural gas prices. \textcite{lv2013modeling} used a variety of linear and nonlinear \ac{GARCH} type models, including models that capture asymmetric effects. Moreover, they used two different distributional assumptions: normality assumption and skewed-student-t. The latter consistently outperformed the normality assumption, regardless of the chosen model. This confirms prior research stating that asset returns often feature heavy tails. Hence using the normal distribution is inappropriate in most cases \autocite{engle2007good}. However, the results did not indicate asymmetric effects. \textcite{chkili2014volatility} also considered asymmetries and long memory effects in the volatility process of natural gas prices and other commodities. They found asymmetric effects in all examined commodities, which contrasts the results of \textcite{lv2013modeling}. \textcite{hailemariam2019drives} investigated how supply and demand shocks affect natural gas price volatility. They found that the effects greatly depend on the market regime and that several structural breaks influenced the volatility response. The global gas market liberalization altered the gas price volatility in particular.


This paper contributes to the literature by creating probabilistic price forecasting models for natural gas Day-Ahead and Month-Ahead products. We propose state-space time series models assuming a heavy-tailed error distribution. The model development builds on a rigorous descriptive data analysis and results from previous research. The proposed models consider many aforementioned and novel effects and significantly outperform popular models presented in the literature. We provide an extensive discussion and validation of each model component and a detailed analysis of the probabilistic predictive performance.


The remainder of this paper is structured as follows. \Cref{data} presents an overview of the European gas market, the considered price data, and many stylized facts. The proposed models are presented and discussed in \Cref{models}. \Cref{forecastingstudy} contains the forecasting study design and benchmark models. \Cref{results} presents extensive model diagnostics and discusses the predictive performance. \Cref{conclusion} concludes.

\section{Market Data and Stylized Facts}\label{data}

\begin{figure}[ht]
  \centering
  \fbox{\includegraphics[width=0.975\columnwidth]{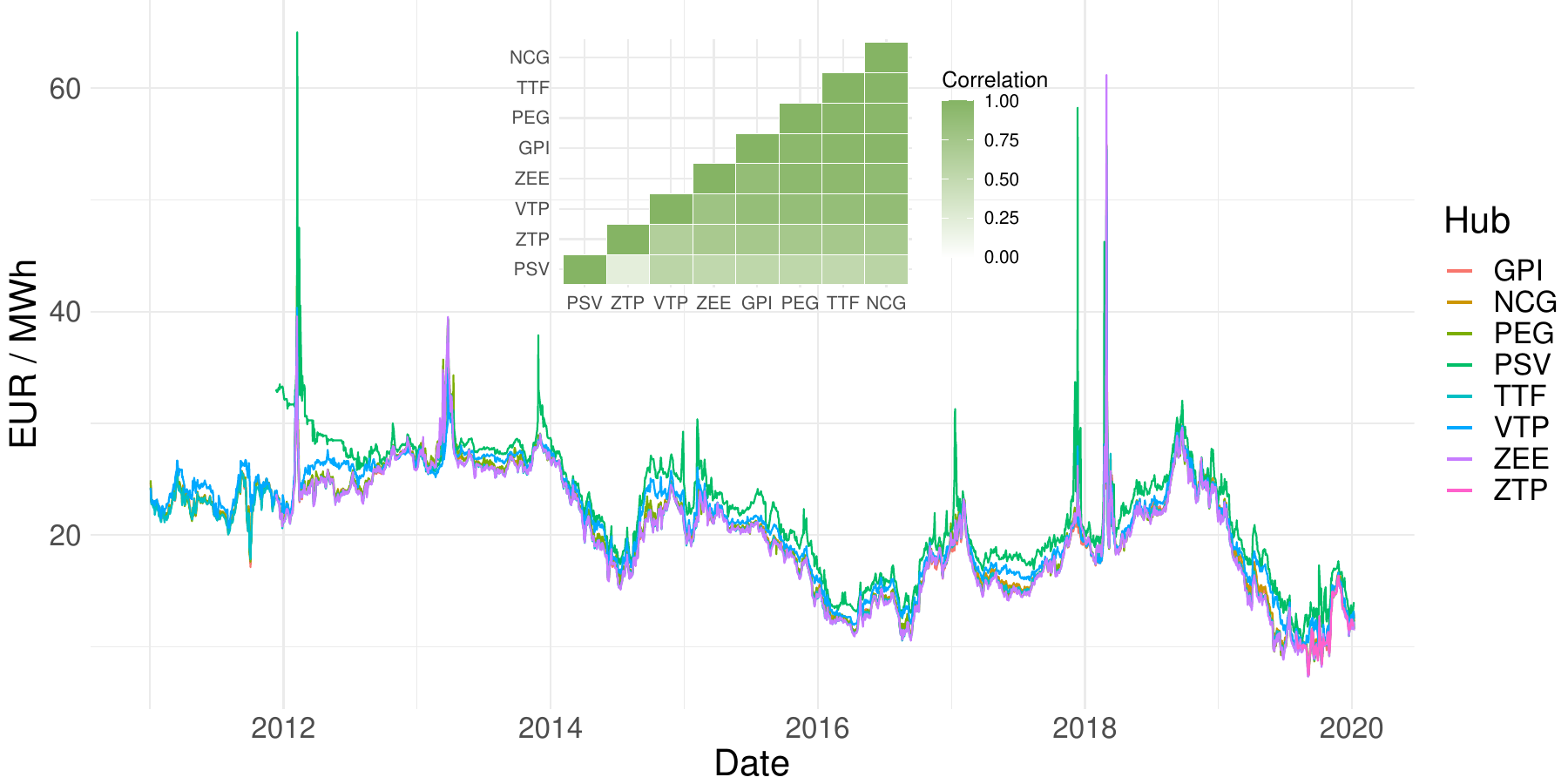}}
  \caption{Natural gas Day-Ahead prices at different European trading points and their sample correlation}
  \label{fig:ts_hubs_da}
\end{figure}

We use daily closing prices on future markets for natural gas in Europe. We consider the two most liquid products: the Day-Ahead and Month-Ahead product. The choice of the trading hub does not matter much since European natural gas markets are integrated to a great extent; therefore, prices are highly correlated. \Cref{fig:ts_hubs_da} presents this correlation for the Day-Ahead time series. That is, we chose pricing data from the Title Transfer Facility (TTF) Hub since it provides sufficient historical data and is rated the most important trading hub in Europe \footnote{Review of the Gas Hub Assessment, European Federation of Energy Traders (EFET), 2019}. Further, a detailed discussion about European natural gas market integration can be found in \textcite{hamie2021modeling}.

\begin{figure}[h]
  \centering
  \fbox{\includegraphics[width=0.975\columnwidth]{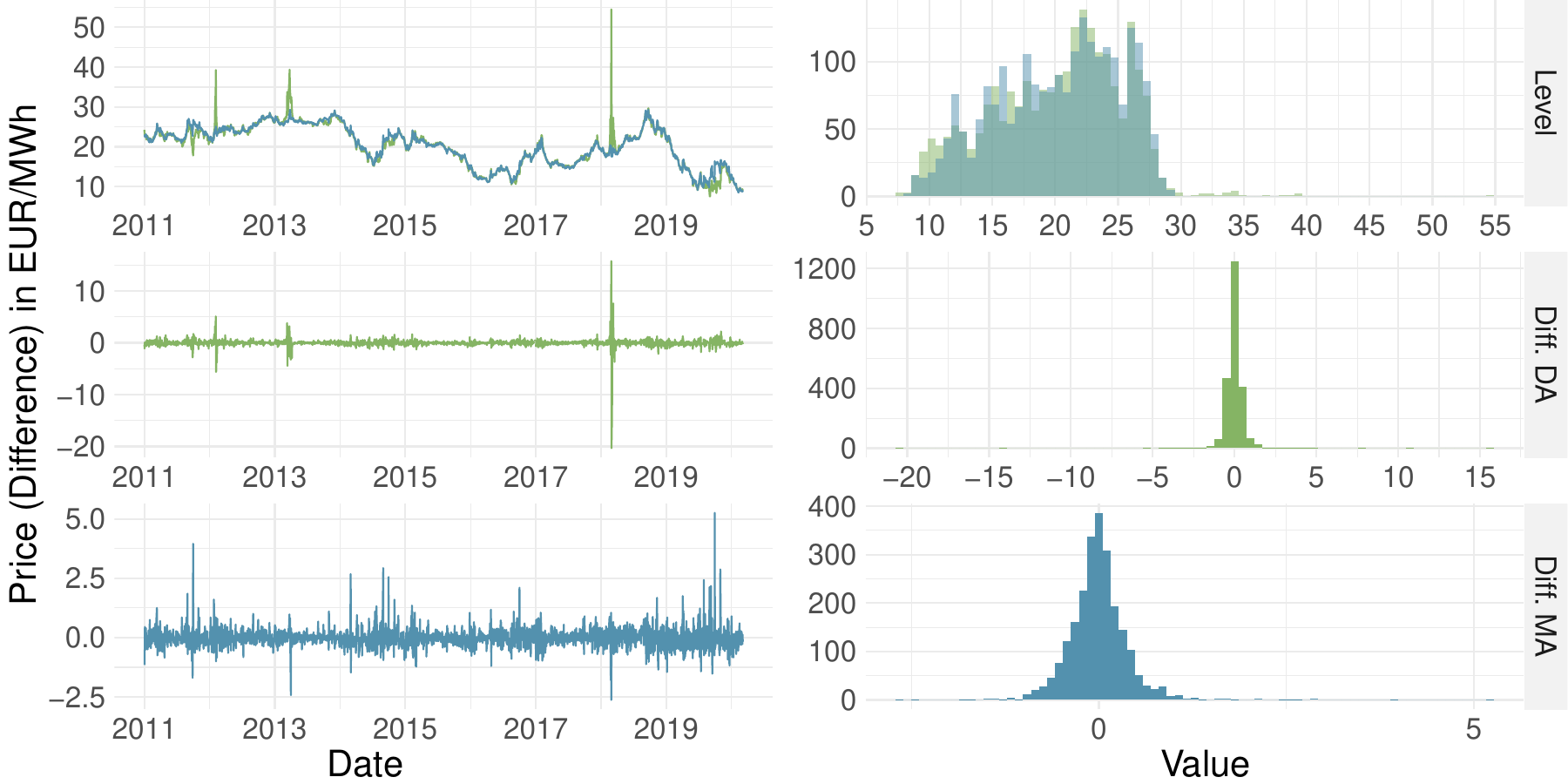}}
  \caption{Original and differenced Day-Ahead (green) and Month-Ahead (blue) time series of natural gas prices (left) with corresponding histograms (right)}
  \label{fig:ts}
\end{figure}

\Cref{fig:ts} presents an overview of both analyzed time series. The data ranges from January 2011 to March 2020. The time series of the Day-Ahead product (green) features some large upward spikes, which lead to a heavy-tailed return distribution. Those spikes are likely caused by unexpected high demand. This pattern is not present in the Month-Ahead product since it is not suitable for offsetting these unexpected short-term shocks. We also observe volatility clusters in both processes, indicating that the conditional volatility is not constant. There is neither an apparent trend nor a cycle present in the data.

We fitted a generalized student-t distribution to the differenced time series to obtain a tail index estimate to examine the heavy tails further. We denote the estimates by $\tau$. They are presented in \Cref{sum_stats_ts} along with other descriptive statistics and the p-values of an \ac{ADF} and a \ac{PP} unit root test. The tail-index estimate of the Day-Ahead time series is particularly low, with a value merely larger than 2. This indicates that the second moment hardly exists. Therefore standard least-squares methods likely lead to unreliable results due to the considerable influence of particular observations. This underlines that the normality assumption does not hold.

\begin{table}[h] \centering
  \caption{Descriptive statistics of the Day-Ahead time series, $\tau$ denotes the degrees of freedom from a fitted generalized student-t distribution, PP and ADF denote the P-Value of the Philipps-Perron and Augmented Dickey-Fuller unit root test respectively}
  \label{sum_stats_ts}
  \resizebox{1\textwidth}{!}{%
    \begin{tabular}{@{\extracolsep{5pt}}lrrrrrrrrrr}
      \\[-1.8ex]\hline
      \hline                                                                                                                                                                                                                                                       \\[-1.8ex]
      Statistic              & \multicolumn{1}{c}{Mean} & \multicolumn{1}{c}{St.Dev.} & \multicolumn{1}{c}{Min} & \multicolumn{1}{c}{Pctl(25)} & \multicolumn{1}{c}{Median} & \multicolumn{1}{c}{Pctl(75)} & \multicolumn{1}{c}{Max} & $\tau$ & PP      & ADF      \\
      \hline                                                                                                                                                                                                                                                       \\[-1.8ex]
      Level (DA)             & 20.15                    & 5.25                        & 7.57                    & 16.25                        & 20.99                      & 24.00                        & 54.52                   & -      & $<0.01$ & 0.18     \\
      First Differences (DA) & $-$0.01                  & 0.86                        & $-$20.31                & $-$0.23                      & $-$0.01                    & 0.21                         & 15.77                   & 2.26   & $<0.01$ & $< 0.01$ \\
      Level (MA)             & 20.25                    & 4.92                        & 8.53                    & 16.25                        & 21.00                      & 24.25                        & 29.35                   & -      & 0.48    & 0.52     \\
      First Differences (MA) & $-$0.01                  & 0.42                        & $-$2.62                 & $-$0.21                      & $-$0.02                    & 0.18                         & 5.25                    & 2.86   & $<0.01$ & $<0.01$  \\
      \hline                                                                                                                                                                                                                                                       \\[-1.8ex]
    \end{tabular}
  }
\end{table}

\begin{figure}[h]
  \centering
  \fbox{\includegraphics[width=0.975\columnwidth]{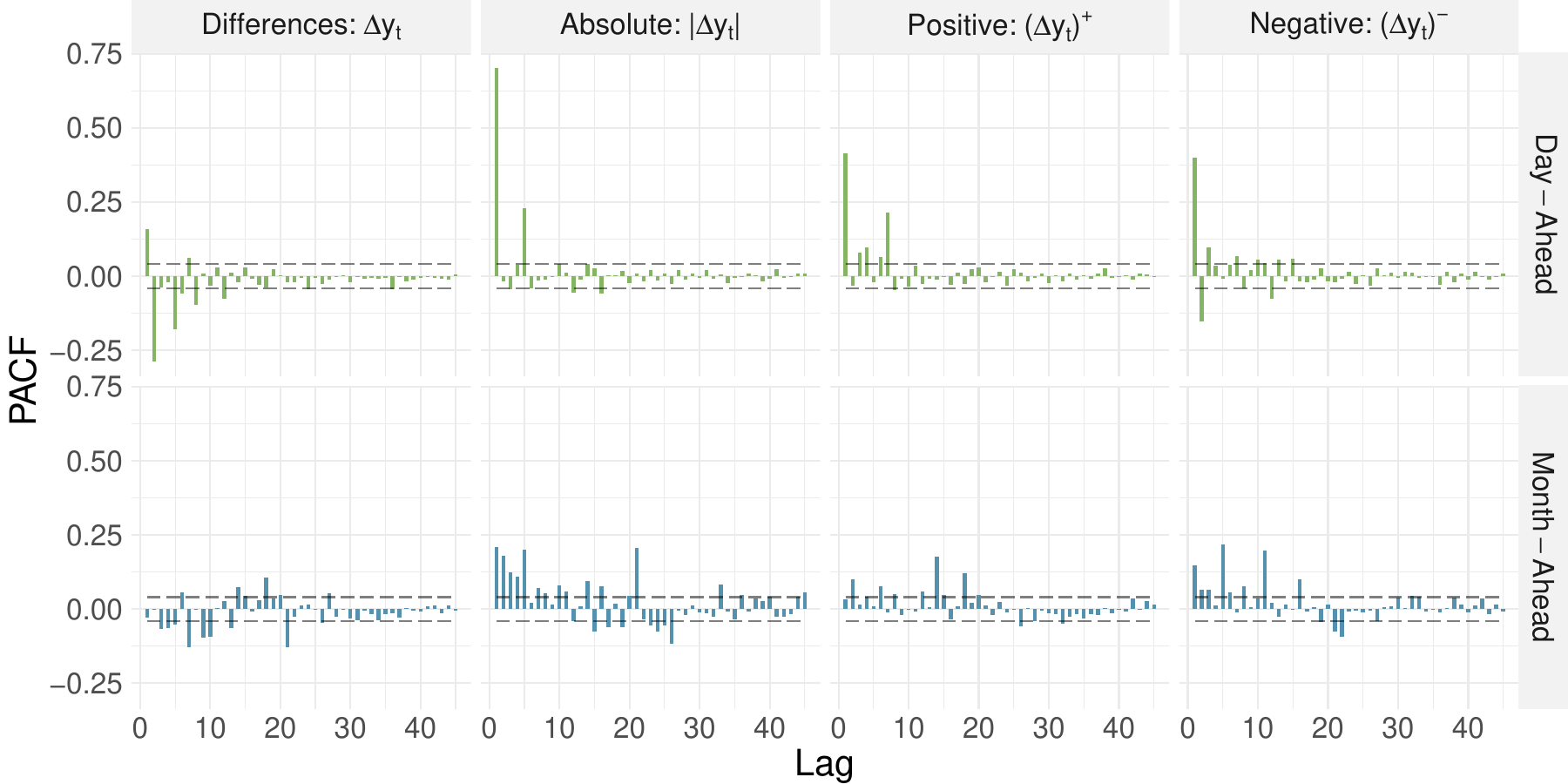}}
  \caption{\label{fig:pacf}Sample PACF of the differenced Day-Ahead (green) and Month-Ahead time series (blue) and transformations thereof up to Lag 40 with 5\% confidence intervals}
\end{figure}

\Cref{fig:pacf} shows the sample \ac{PACF} of both differenced time series, as well as the PACF of their absolute, positive and negative parts. The PACF of the differenced Day-Ahead time series features some significance at lower lags, while the differenced Month-Ahead time series has significance around lag 21. The monthly rollover of the product likely causes the latter. This pattern is also apparent in the sample \ac{PACF} of the absolute differences. Thus, the rollover likely influences the mean and the conditional variance of the Month-Ahead prices. The \ac{PACF} of the absolute differences of the Day-ahead and Month-Ahead products are also significantly different from zero at lower lags, indicating short-term ARCH or GARCH effects in the conditional variance.
Comparing the negative and positive parts of the differences reveals some disparity for both products. That disparity likely indicates leverage effects. That is, modeling asymmetric effects in the volatility process probably improves the forecasting performance.

\begin{figure}[h]
  \centering
  \fbox{\includegraphics[width=0.975\columnwidth]{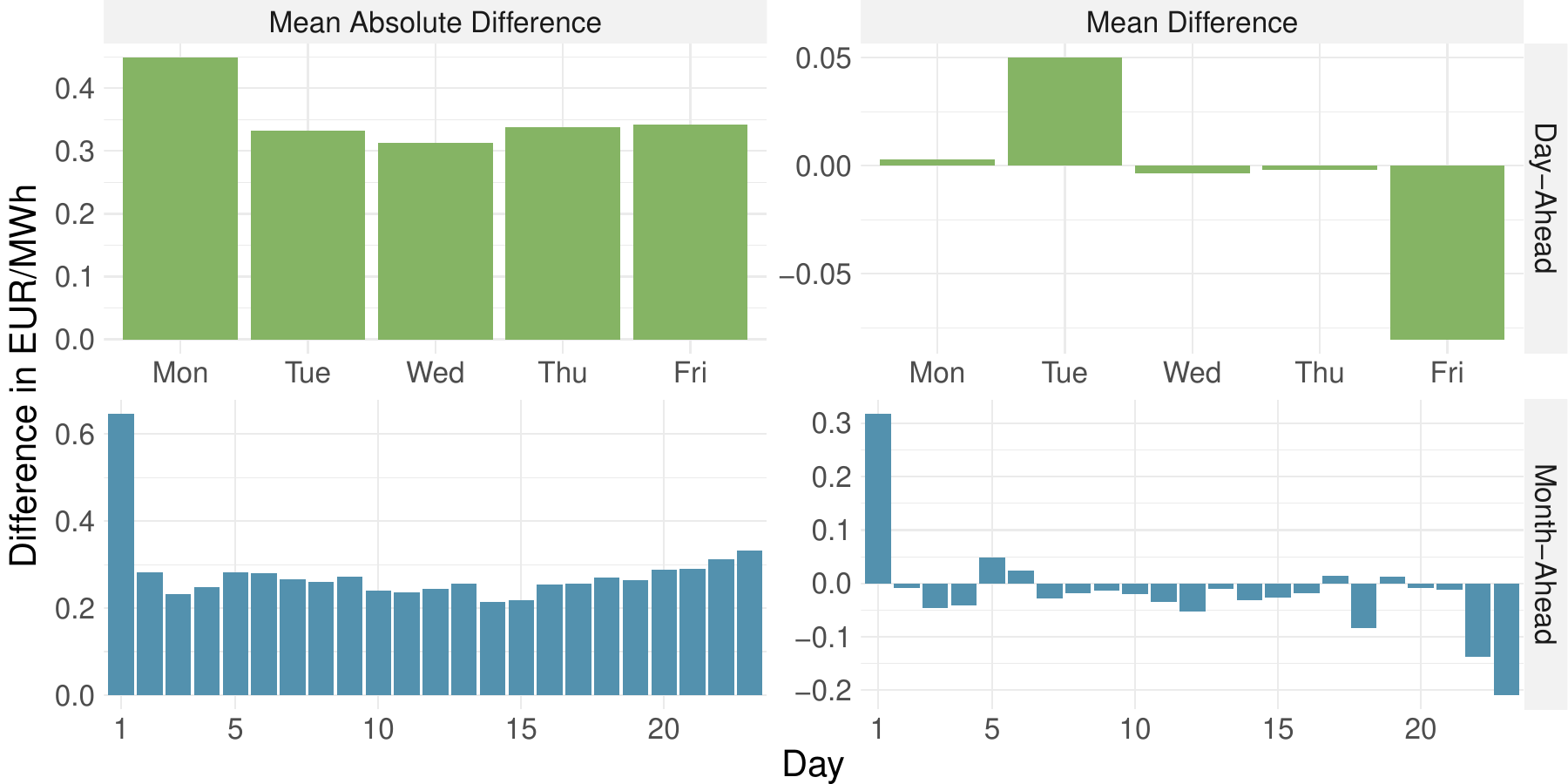}}
  \caption{Mean \textit{absolute} differences (left) and mean differences of the Day-Ahead (green) and Month-Ahead (blue) time series, per weekday (top) and trading day (bottom)}\label{fig:day_fx}
\end{figure}

\Cref{fig:day_fx} depicts the mean absolute difference and mean difference for both products on different weekdays (Day-Ahead) and days of the month (Month-Ahead). The Day-Ahead product features increased absolute differences on Mondays, likely caused by the increased amount of information that emerges over the weekend. The price change from Friday to Monday reflects the uncertainty concerning the whole weekend, which is generally higher than the uncertainty only concerning one day \autocite{mu2007weather}. A suitable volatility model should consider this effect since it will likely improve the probabilistic forecasting performance.
The mean absolute difference of the Month-Ahead product is increased on the first trading day of the month. The latter indicates increased conditional volatility. This is intuitive since the product rollover causes the underlying delivery period to shift by one month. This effect will also be addressed by the proposed Month-Ahead model presented in \Cref{sec:mod_ma}.

However, the positive mean difference on the first trading day of the month is noteworthy. One possible explanation for this pattern is the uncertainty concerning the delivery month. While the uncertainty is lowest on the last day of the month (because it is close to delivery), it is highest on the first of every month. Market participants require risk premia as compensation for such risks. The risk-premia demanded by buyers and sellers usually cancel out. However, if the risk-aversion is distributed asymmetrically, the risk premia would differ between both sides of the market. Thus, the risk premia do not cancel out, and the prices will be affected. In particular, if the supply side is more risk-averse, it requires higher risk premia than the seller side, which leads to a price increase. Therefore, the sudden price increase that we observe on the first day of the month is possibly caused by asymmetric risk aversion. This also implies that the sudden price increase will gradually decay as the uncertainty concerning the delivery month diminishes. Both effects will be considered and discussed in Sections \ref{models} and \ref{results}. \textcite{roncoroni2017hedging} discuss risk hedging strategies in natural gas markets.

\begin{figure}[h]
  \centering
  \fbox{\includegraphics[width=0.975\columnwidth]{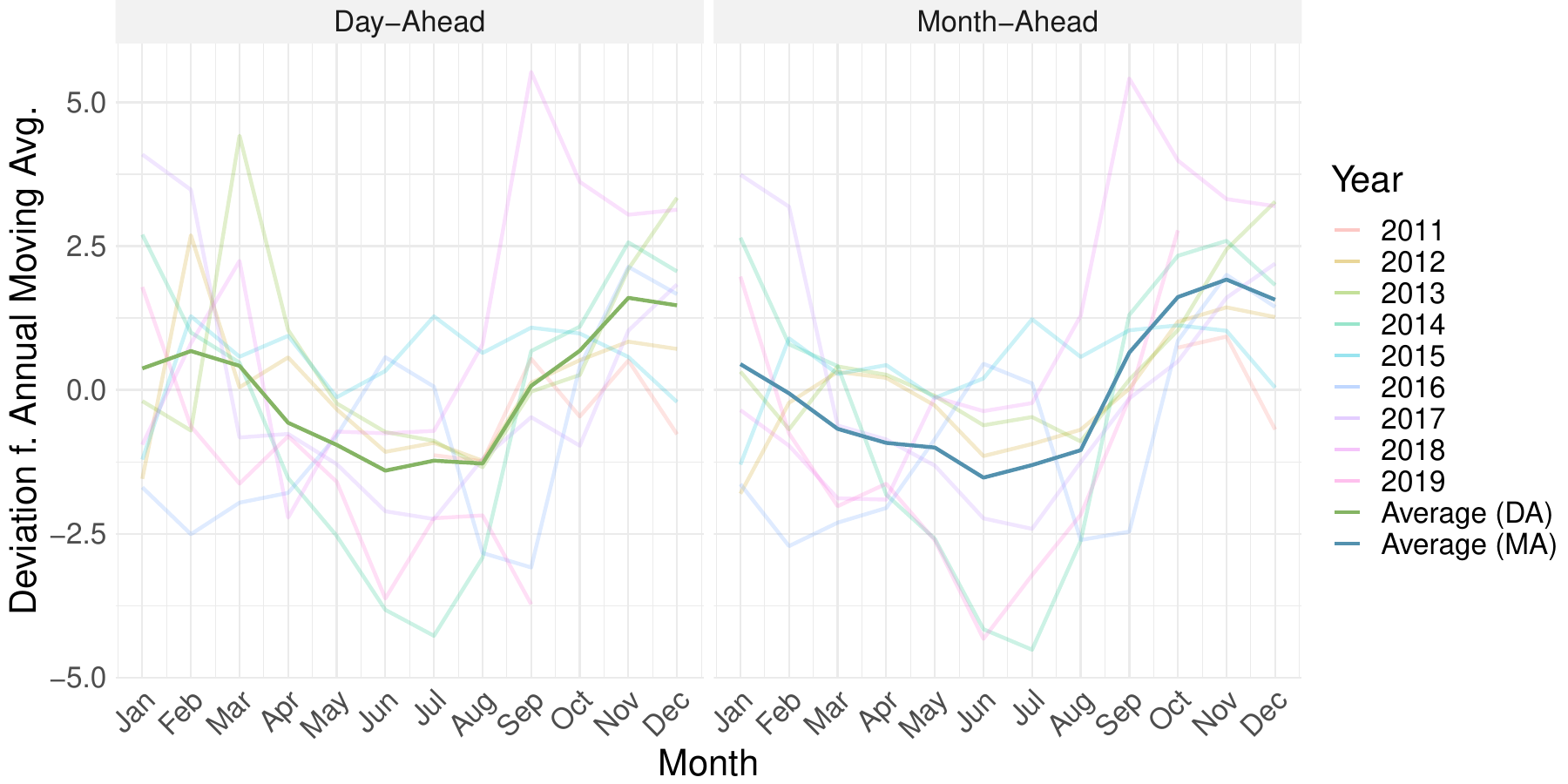}}
  \caption{Monthly deviation from the annual moving average per year (transparent) and averaged over all years (bold, solid) for the Day-Ahead (left) and Month-Ahead (right) time series}
  \label{fig:seasonal}
\end{figure}

\Cref{fig:seasonal} shows the deviation of the monthly averages from the annual rolling average per year (transparent) and the average monthly deviation from the annual average (bold nontransparent). Both time series feature a seasonal pattern with increased prices in the colder winter months. This pattern is likely caused by the increased natural gas demand in winter, caused by heating. The pattern persists because of storage costs which prevent it from being fully diminished by arbitrage transactions \autocite{mirantes2012stochastic, chaton2008some}.

The natural gas price is likely also influenced by various external factors, as discussed in \Cref{introduction}. A natural starting point is considering the prices of related energy products like oil, coal, and power. The literature review indicates that oil \autocite{geng2017relationship} and coal \autocite{li2017analysis} prices influence the natural gas price \autocite{papiez2011analysis}. This is intuitive since coal, oil, and natural gas can be used for power generation. Therefore, they are substitutable to some degree. \textcite{asche2006uk} found an influence of power prices on the natural gas prices. However, the direction of those dependencies may change if the underlying relations change, e.g., caused by political actions.

\begin{figure}[h]
  \centering
  \fbox{\includegraphics[width=0.975\columnwidth]{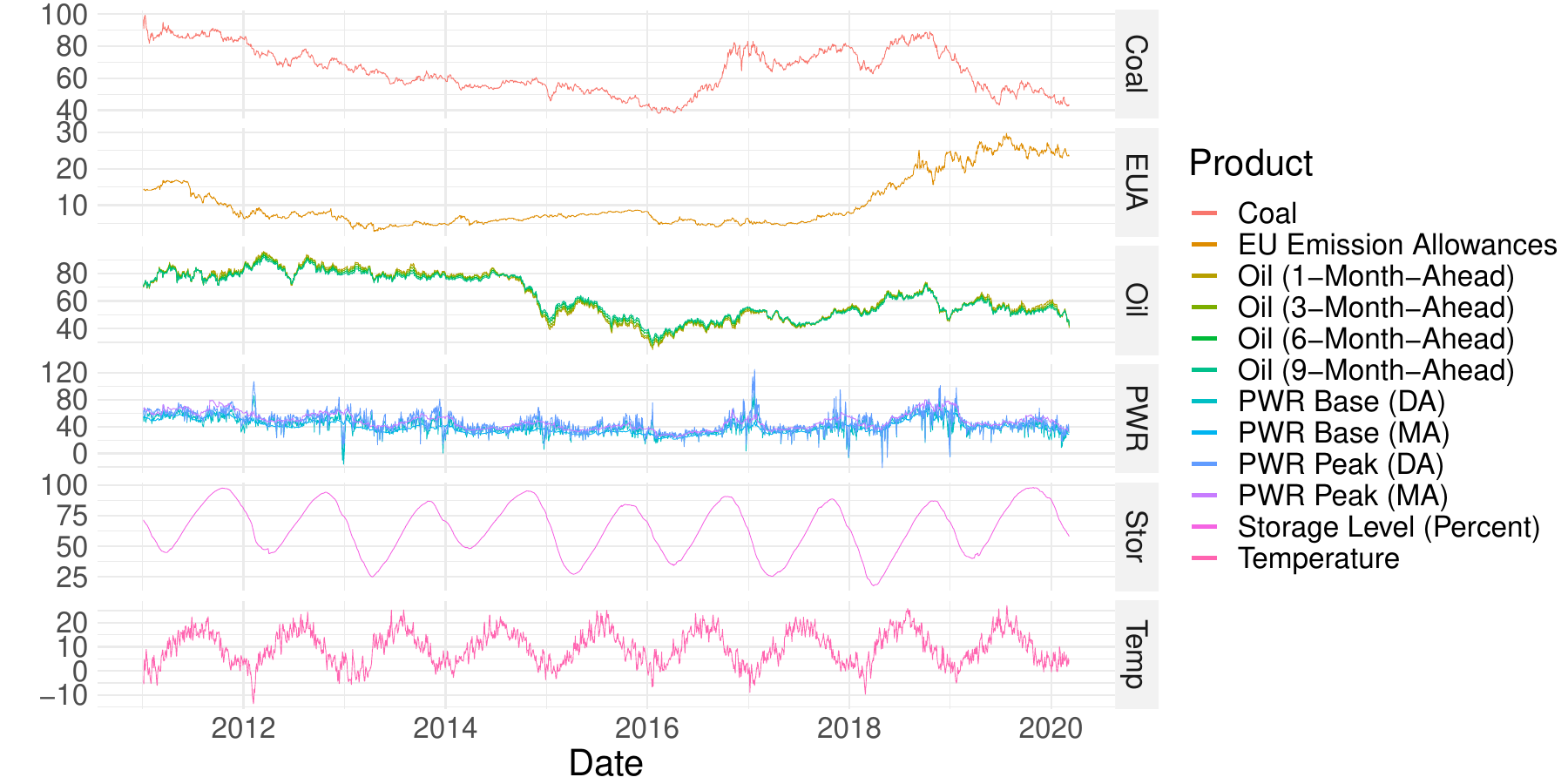}}
  \caption{Time series plot of: Month-Ahead Rotterdam coal price (Coal), EUA price, X-Month-Ahead Brent crude oil prices, Day-Ahead power base and power peak prices of Germany (PWR Base/Peak (DA/MA)), European natural gas storage levels (in percent) and daily averages of the German temperature}\label{fig:ext_regs}
\end{figure}

\Cref{fig:ext_regs} presents the external regressors that we consider. In particular, we take into account the Month-Ahead Rotterdam coal price, the Three-, Six- and Nine-Month-Ahead brent crude oil price, and the German Day-Ahead (DA) and Month-Ahead (MA) power base and power peak prices. Additionally, we consider the daily average Germany temperature, European natural gas storage levels (in percent), and daily spot prices of \ac{EUA}.

For \ac{EUA} the direction in which it influences natural gas prices is unclear. Since using natural gas produces carbon emissions, high \ac{EUA} prices will render the usage of natural gas unattractive. This effect likely leads to a price reduction. On the other hand, high \ac{EUA} prices render the power generation using natural gas more attractive compared to the power generation using coal. However, natural gas has a lower carbon footprint than coal in power production. This substitution effect will lead to increased demand for natural gas due to the fuel switch \autocite{lu2012implications, pratson2013fuel}. Hence, natural gas prices should rise.

However, which of both effects mentioned above dominates has to be estimated. The results are discussed in \Cref{results}.

Other factors that might influence natural gas prices through the influence on consumption are temperature \autocite{karabiber2020forecasting} and storage levels \autocite{chaton2008some}. Both are directly related to the seasonal demand pattern.

\section{Empirical Models}\label{models}

In this section, we discuss the proposed state-space time series models for the Day-Ahead and Month-Ahead product \autocite{Hyndman2008}. Both models are developed in a stepwise procedure to incorporate all effects discussed previously. Only effects that improved the probabilistic forecasting accuracy measured by the out-of-sample \ac{CRPS} were kept. As Day-Ahead and Month-Ahead products feature distinct properties to some extent, we discuss them separately.

\subsection{The Day-Ahead Product Model}\label{sec:mod_da}

The proposed Day-Ahead model is given in equations~\ref{da_level_eq} to~\ref{da_tgarch}:

\begingroup
\begin{subequations}
  \begin{align}
    y_t                  & = \ell_{t-1} + \underbrace{\frac{\psi_1}{5} \sum_{s=t-A-2}^{t-A+2} y_{s}}_{\text{Seasonal}} +  \underbrace{\vphantom{\sum_{s=t-A-2}^{t-A+2}}\zeta_1 \text{Coal}_{t-1}}_{\text{Coal}} + \underbrace{\vphantom{\sum_{s=t-A-2}^{t-A+2}}\zeta_2 \text{EUA}_{t-1}}_{\text{EUA}} + \underbrace{\vphantom{\sum_{s=t-A-2}^{t-A+2}} \zeta_3 \text{PWR}_{t-1}}_{\text{Power}} + d_t \label{da_level_eq} \\
    \ell_{t}             & = \underbrace{\ell_{t-1} + \lambda d_t}_{\text{Exp. Smoothing}} \label{da_state_eq}                                                                                                                                                                                                                                                                                                           \\
    d_t                  & = \underbrace{\sum_{i=1}^{p} \varphi_i d_{t-i} + \sum_{i=1}^{q} \theta_i \varepsilon_{t-i}}_{\text{ARMA Errors}} + \varepsilon_t,                                                                                                                                                                                                                                                             \\
    \varepsilon_t        & \sim \text{SST}(0, \sigma_t, \nu, \tau) \tag{1d}                                                                                                                                                                                                                                                                                                                                              \\
    \sigma_t             & = \begin{cases} \delta \widetilde{\sigma}_t  & \qquad \qquad \qquad \text{Monday} \\ \widetilde{\sigma}_t & \qquad \qquad \qquad \text{Else} \end{cases} \tag{1e}                                                                                                                                                                                                                                                                                                                                                         \\
    \widetilde{\sigma}_t & = \underbrace{\underbrace{\omega + \alpha |\varepsilon_{t-1}| + \beta \sigma_{t-1}}_{\text{Absolute Value GARCH}} + \underbrace{\gamma |\varepsilon_{t-1}| \mathbb{1}(\varepsilon_{t-1} < 0)}_{\text{Leverage}}}_{\text{TGARCH}} \tag{1f} \label{da_tgarch} .
  \end{align}
\end{subequations}
\endgroup

The model exploits the autoregressive structure with an exponential smoothing process (\ref{da_level_eq} and~\ref{da_state_eq}) with \ac{ARMA} errors denoted by $d_t$~\autocite{de2011forecasting}. The price at time $t$ is denoted by $y_t$ and $\ell_t$ denotes the level component of the state space model. The seasonal structure is implemented using a one-week moving average, which is lagged by one year. One week usually contains five observations since the Day-Ahead product is only traded on working days. We also considered using seasonal lags or trigonometric Fourier terms to incorporate seasonality. However, the averaged lag proved superior. Moreover, we extend the level equation (\ref{da_level_eq}) of the model by the coal price (coal), the price of \ac{EUA}, and the Day-Ahead peak power price (PWR).

The volatility process was modeled using a \ac{TGARCH} process \autocite{bollerslev2010volatility}. Compared to the traditional \ac{GARCH} model, it reduces the influence of large errors by modeling the conditional standard deviation instead of the conditional variance. Thereby, the conditional standard deviation depends on a constant, its own lagged values up to lag $p$, the lagged absolute errors up to lag $q$, and a leverage effect for negative errors. Refraining from squaring the errors is particularly relevant for natural gas prices due to the heavy-tails, which we discussed in \Cref{data}.

We address the increased volatility on Mondays by leveraging the estimated standard deviation on Mondays. This approach is necessary to avoid increasing the variance on other weekdays as well. The latter would be the case when using a Monday dummy (due to the recursive structure of the \ac{TGARCH} process).

We assume the errors to be skewed-student-t distributed, i.e. $\varepsilon_t \sim \text{SST}$. The skewed version of the generalized Student-t distribution was developed by \textcite{fernandez1998bayesian}, who proposed a general approach for introducing skewness in symmetric distributions and applied that approach to the generalized Student-t distribution. This distribution was later reparameterized by \textcite{wurtz2006parameter} such that $\mu$ denotes the mean and $\sigma$ denotes the standard deviation. The density is defined as follows:
\begin{align}
  f_{\mu, \sigma, \nu, \tau}(y) & = \begin{cases}
    \frac{c}{\sigma_0}\big[1+\frac{\nu^2 y^2}{\tau}\big]^{-(\tau+1)/2} & \text{if $y < \mu_0$}    \\
    \frac{c}{\sigma_0}\big[1+\frac{y^2}{\nu^2 \tau}\big]^{-(\tau+1)/2} & \text{if $y \geq \mu_0$}
  \end{cases} \label{sstpdf}                            \\
  \text{with:}
  \quad \mu_0                   & = \mu - \sigma m /s \nonumber                                          \\
  \quad \sigma_0                & = \sigma / s \nonumber                                                 \\
  c                             & = 2\nu[(1+\nu^2)B(1/2,\tau/2)\tau^{1/2}]^{-1} \nonumber                \\
  m                             & = \frac{2\tau^{1/2}(\nu - \nu^{-1})}{(\tau -1)B(1/2, \tau/2)}\nonumber \\
  s                             & = \sqrt{\frac{\tau}{\tau-2}(\nu^2+\nu^{-2}-1)- m^2}\nonumber
\end{align}
where $y, \mu \in \mathbb{R}$, $\sigma, \nu > 0$, and $\tau >2$. Here, $\nu$ denote the skewness parameter and $\tau$ denotes the degrees of freedom. The distribution is symmetric if $\nu = 1$.

This distribution features heavier tails than the Gaussian distribution and is, therefore, better suited for this data. Moreover, it allows for skewness, which was also discussed in related literature and \Cref{data}. Finally, we discuss the importance of each model component in detail in \Cref{results}.

\subsection{The Month-Ahead Product Model}\label{sec:mod_ma}

The Month-Ahead model considers all characteristics that were discovered in \Cref{data}. Preliminary analysis showed that complex autoregressive models do not outperform simple random walk models. Hence, we extended a random walk model to incorporate all the characteristics of the product. The proposed Month-Ahead model is given in equations~\ref{level_eq} to~\ref{ma_tgarch}.

\begingroup
\begin{subequations}
  \begin{align}
    y_t                  & = \varphi_0 + \underbrace{\Phi_t}_{\substack{\text{Risk }                                                                                                                                                                                          \\ \text{Rollover}}} + \underbrace{\psi_1 \Bar{y}_{t-1Y}^M}_{\text{Seasonal}} + \underbrace{\zeta_1 \text{EUA}_{t-1}}_{\text{EUA}} + \underbrace{\zeta_2 \text{Oil}_{t-1}}_{\text{Oil}} + \underbrace{\zeta_3 \widetilde{T}_{t-1}}_{\text{Temperature}} + \varepsilon_{t} \label{level_eq} \\
    \Phi_t               & = \begin{cases} \varphi_1 y_{t-1, \text{2MA}} & \quad \text{First Trading Day per Month}                                                                                                                    \\[1ex]
              \smash[b]{%
              \underbrace{\vphantom{\eta \frac{D_t^{\text{LTD}}}{D_t^{\text{M}}}}\varphi_1 y_{t-1}}_{\text{Rollover}} + \underbrace{\eta \frac{D_t^{\text{LTD}}}{D_t^{\text{M}}}}_{\substack{\text{Risk } \\ \text{Deduction}}}} & \quad \text{Else}
    \end{cases}                                                                                                                                                                                                                       \\[8ex]
    \varepsilon_t        & \sim \text{SST}(0, \sigma_t, \nu, \tau)                                                                                                                                                                                                            \\[2ex]
    \sigma_t             & = \begin{cases} \delta \widetilde{\sigma}_t  & \qquad \qquad \qquad \qquad \text{First Trading Day per Month} \\ \widetilde{\sigma}_t & \qquad \qquad \qquad \qquad \text{Else} \end{cases}                                                                                                                                                                                                                       \\
    \widetilde{\sigma}_t & = \underbrace{\underbrace{\omega + \alpha |\varepsilon_{t-1}| + \beta \sigma_{t-1}}_{\text{Absolute Value GARCH}} + \underbrace{\gamma |\varepsilon_{t-1}| \mathbb{1}(\varepsilon_{t-1} < 0)}_{\text{Leverage}}}_{\text{TGARCH}} \label{ma_tgarch}
  \end{align}
\end{subequations}
\endgroup

We modeled the product rollover by using the most recent price of the Two-Month-Ahead product $y_{t-1, \text{2MA}}$ as a predictor for the Month-Ahead product on the first trading day of each month. The model considers the most recent price of the Month-Ahead product $y_{t-1}$ on all other days. Additionally, we added a risk component. The buildup of the risk premia does not need to be modeled since the approach mentioned above to address the model rollover accounts for it already. However, the risk premia gradually decay over the month. We assume this decay to be linear. That is, we weight the decay by the number of days since the last trading day $D_t^{\text{LTD}}$ to account for a more significant risk reduction, e.g., on Mondays. We also normalize this component using the total number of days in a given month $D_t^{\text{M}}$.
Moreover, we included a seasonal component. We used the monthly average lagged by one year $\Bar{y}_{t-1Y}^M$. This approach has a clear advantage compared to, e.g., Fourier terms: it accounts for the monthly structure of the product, i.e., the rollover.

The mean process was further extended by the lagged price of European Emission Allowances, the lagged price of Oil, and the lagged and smoothed temperature denoted by $\text{EUA}_{t-1}$, $\text{Oil}_{t-1}$ and $\widetilde{T}_{t-1}$ respectively. The smoothed temperature  is defined as $\widetilde{T_{t}} = 0.95 \widetilde{T}_{t-1} + (0.05) T_{t}$. That is, we follow \textcite{gaillard2016additive} by smoothing the temperature exponentially with a persistence parameter of $0.95$.
Lastly, the model was supplemented by a TGARCH process similar to the process of the Day-Ahead model. However, the conditional standard deviation is elevated on the first trading day of every month instead of every Monday.
We assume that the errors are SST distributed as in the Day-Ahead model.

\section{Forecasting Study}\label{forecastingstudy}

\subsection{Forecasting study design and Benchmarks}
The validity and forecasting performance of the models was evaluated using one-step-ahead out-of-sample forecasts obtained
by an expanding window forecasting study.

That is, the estimation window expands as the forecasting study progresses. We chose this approach to utilize an increasing number of observations for estimation. This approach is more suitable than a rolling window forecasting study due to the heavy-tailed data. We initialized the calibration window with four years of observations (exactly 1012/1012 days Day-Ahead/Month-Ahead), leaving more than four years for validation (exactly 1056/1079 days Day-Ahead/Month-Ahead). The models were estimated using maximum likelihood.

Moreover, we set up three different benchmark models from recent literature. Since all three models were only used to do point-forecasting, we had to set a distributional assumption ourselves. We chose to estimate each model with the common normality assumption and the T assumption, which better fits this data, as discussed above.

First, we replicated the time series benchmark model of \textcite{siddiqui2019predicting}. They used an \ac{ARIMA}(2,1,2) model. Since they did not specify how this model was calibrated, we just chose the same order.
Second, we replicated the Random Forest model of \textcite{herrera2019long} which outperformed all other considered models by a great margin. However, we had to adapt the hyperparameters here since \citeauthor{herrera2019long} applied this model to monthly data. That is, we fitted 500 trees like \citeauthor{herrera2019long}, created 270 lag variables which correspond to roughly one year of observations, and considered 23 ($\sqrt{270\times2} \approx 23$) variables at each split in contrast to 7.
Lastly, we fit a vector autoregressive model like \textcite{geng2017relationship} to the differenced time series of natural gas and Month-Ahead oil prices. We determined the \ac{VAR} order by using the Akaike information criterion.

We evaluated the benchmarks using the same forecasting study design as described above. We estimated the missing distributional parameters by fitting a Normal and Student-T-distribution to the error terms. These parameters were also estimated using maximum likelihood. Thus, we receive $2\times3=6$ benchmark models in total.

\subsection{Evaluation Measures}

We use four scoring functions to evaluate the forecasting performance on the aforementioned out-of-sample forecasts. Namely, the \ac{MAE}, \ac{RMSE}, \ac{CRPS}, and Pinball Loss. \Cref{tab:scoring} presents these scoring functions formally. While \ac{MAE} and \ac{RMSE} are very popular, they do not evaluate the whole probabilistic forecast. The (R)MSE should be used only to evaluate mean forecasts. Similarly, the \ac{MAE} is appropriate only for evaluating median forecasts~\autocite{gneiting2011making}.
\begin{table}[h!]
  \caption{\label{tab:scoring}Scoring functions used for evaluating the forecasting performance of mean ($\hat{\mu}_t$), median ($\tilde{\mu}_t$), $p$-quantile ($\widehat{Q}_{t}(p)$) or distribution ($\widehat{F}_{t}$) predictions and corresponding observations $y_t$.}
  \small
  \centering
  \begin{tabular}[t]{ll}
    \toprule
    \textbf{Name}                       & \textbf{\text{Scoring Function}}                                                                                                       \\
    \midrule
    Root Mean Squared Error             & \(\displaystyle \text{RMSE}(y_t, \hat{\mu}_t) = \sqrt{\frac{1}{T}\sum_{t=1}^T {(y_t - \hat{\mu}_t)}^2}\)                               \\
    \midrule
    Mean Absolute Error                 & \(\displaystyle \text{MAE}(y_t, \tilde{\mu}_t) = \frac{1}{T}\sum_{t=1}^{T} |y_t - \tilde{\mu}_t |\)                                    \\
    \midrule
    Pinball Loss for probability $p$    & \( \displaystyle \text{PL}(\widehat{Q}_{t}(p), y_t, p) = \begin{cases}
      p (y_t-\widehat{Q}_{t}(p))      & y_t \geq \widehat{Q}_{t}(p) \\
      (1-p)(\widehat{Q}_{t}(p) - y_t) & y_t < \widehat{Q}_{t}(p)
    \end{cases} \)                                                 \\
    \midrule
    Continious Ranked Probability Score & \(\displaystyle \text{CRPS}(\widehat{F}_{t}(z), y_t) = \int_{\mathbb{R}} {(\widehat{F}_{t} (x) - \mathbb{1}( z > y_t ))}^2 dx       \) \\
    \bottomrule
  \end{tabular}
\end{table}
Forecasting the whole probabilistic distribution yields a significant advantage: The possibility to obtain several predictors like the mean, the median, or other quantiles of interest without the need to specify the desired quantity of interest beforehand. This also allows us to use multiple scoring functions appropriately by evaluating the respective quantities of the distribution. However, as discussed above, \ac{MAE} and \ac{RMSE} are only suitable for evaluating the mean and median. In contrast, \ac{CRPS} and Pinball Loss can be used for evaluating the full predictive distribution. Both are also strictly proper. This means that they reach a unique minimum when the probabilistic forecast equals the actual probability distribution. Hence they will identify the true model if it is considered \autocite{gneiting2007strictly}. The Pinball loss evaluates the performance of individual quantile forecasts. One can calculate the Pinball loss of a probabilistic forecast for arbitrary probabilities of choice using the quantile function. In this paper, the Pinball loss is calculated on a probability grid containing all percentiles. The results are then plotted as a line graph, usually creating a bell shape. As with the other loss functions, the model performance is better the smaller the Pinball Loss is.

The Pinball Loss is directly related to the \ac{MAE} and the \ac{CRPS}. Calculating the Pinball Loss at the 50\% probability yields the \ac{MAE}/2. Further we can use the Pinball Loss to approximate the \ac{CRPS}:
\begin{align*}
  \text{CRPS}(\widehat{F}_{t}(z), y_t) = 2 \int_0^{1}  \text{PL}(\widehat{Q}_{t}(p), y_t, p)  dp.
\end{align*}
That is, evaluating the Pinball Loss on a dense grid approximates the CRPS \autocite{nowotarski2018recent}.
Both measures have their advantages. The \ac{CRPS}, on the one hand, summarizes the probabilistic forecasting performance in one single scalar but does not offer insights into the forecasting performance of particular quantile forecasts. The Pinball Loss, on the other hand,  resolves the forecasting performance in more detail. However, presenting Pinball losses over a grid with high resolution requires graphical methods almost surely.

Additionally, we use the \ac{DM} for model comparison \autocite{diebold2002comparing}.
The test statistic is defined as follows:
\begin{align}
  t_{DM} &
  = \frac{\overline{\Delta}_{A,B}}{\sigma\left(\overline{\Delta}_{A,B}\right)}                  
  \qquad \qquad \text{with} \qquad \qquad   
  \overline{\Delta}_{A,B}             = \frac{1}{N} \sum_{i = 1}^N \left( L(\varepsilon_{A,i}) -L(\varepsilon_{B,i}) \right) 
\end{align}

where \( \overline{\Delta}_{A,B} \) denotes the loss differential of the two competing models \( A \) and \( B \). \( L \) denotes the loss function which depends on the forecasting error \( \varepsilon_{A,i} \) and \( \varepsilon_{A,i} \) respectively. The Null hypothesis of this test states that the loss differential \( \overline{\Delta}_{A,B} \) is not different from zero. The \ac{DM} is applied using the loss functions presented \Cref{tab:scoring}.

\section{Results and Discussion}\label{results}

\subsection{Model Diagnostics}

\begin{figure}[h!]
  \centering
  \fbox{\includegraphics[width=0.975\columnwidth]{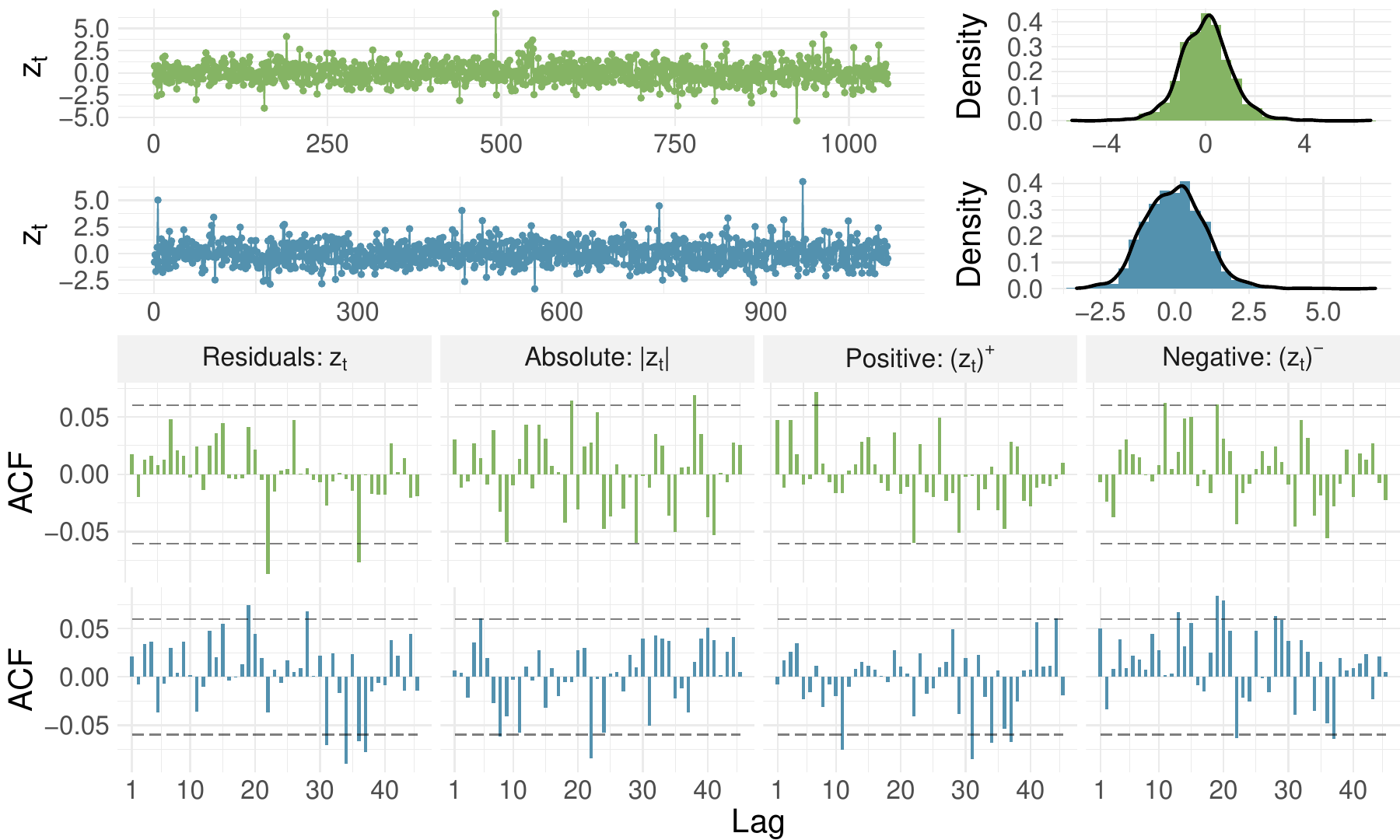}}
  \caption{Standardized residuals of the proposed Day-Ahead (top-left, green) and Month-Ahead (top-left, blue) model and corresponding densities (top-right) as well as sample autocorrelations (ACF) up to lag 40 of the Day-Ahead (green) and Month-Ahead (blue) standardized residuals and transformations thereof, 5\% confidence intervals are indicated by grey dotted lines}
  \label{fig:rp_da_ma}
\end{figure}

We used several measures to assess the model validity before evaluating the predictive performance. First, we visually inspected the standardized mean residuals of both models. These are presented in \Cref{fig:rp_da_ma}. A comparison to \Cref{fig:pacf} shows that both models exploit the majority of the autocorrelation in the data. The line graph and the histogram do not reveal any striking patterns either.

We conducted the \textit{energy test of independence} to test for independence of the standardized mean residuals and several selected quantities \autocite{szekely2007measuring, szekely2013energy}. We denote the standardized residuals by $z_t=\varepsilon_{t} / \sigma_{t}$. This test is based on the distance correlation $\mathcal{R}$, which only requires finite first moments to exist. Distance correlation measures any correlation between two random variables in arbitrary dimensions. That is, $\mathcal{R} = 0$ only holds if X and Y are independent. This is used to create the \textit{energy test of independence}. The null hypothesis of this test is \textit{independence}: $\mathcal{H}_0: y_t \!\perp\!\!\!\perp z_t$. \Cref{tab:da_ma_independence} presents the test results for the standardized residuals of the Day-Ahead and Month-Ahead models. In the Day-Ahead case,  independence is rejected for the dependent variable $y_t$. This indicates that there are still unexploited dependencies. However, these may be nonlinear and potentially weak, which makes exploiting them difficult. The latter also holds for the coal price, the price of \ac{EUA} and the seasonal component, although the P-Values are much higher for these quantities.
For the Month-Ahead model, independence is not rejected for all variables. However, the P-Value of the dependent variable is close to rejection which may indicate some residual dependencies.

\begin{table}[t!]

  \caption{\label{tab:da_ma_independence}Test statistics and p-values of the Energy test of Independence conducted to test whether $z_t$ and the denoted listed are independent}
  \centering
  \begin{tabular}[t]{llrr}
    \toprule
    \textbf{Description         } & \textbf{Variable                } & \textbf{Statistic} & \textbf{P-Value} \\
    \midrule
    \multicolumn{4}{@{}l}{\textbf{\DA{Day-Ahead}}}                                                            \\
    Day-Ahead Price               & $y_t$                             & 20.147             & 0.010            \\
    Seasonal                      & $\sum_{s=t-A-2}^{t-A+2} y_{s}$    & 9.686              & 0.050            \\
    Coal price lagged             & $\text{Coal}_{t-1}$               & 41.122             & 0.030            \\
    EUA price lagged              & $\text{EUA}_{t-1}$                & 30.279             & 0.010            \\
    Power Price lagged            & $\text{PWR}_{t-1}$                & 17.147             & 0.515            \\
    Monday Dummy                  & Monday                            & 0.379              & 0.337            \\
    Standardized residuals lagged & $z_{t-1}$                         & 1.240              & 0.485            \\

    \midrule
    \multicolumn{4}{@{}l}{\textbf{\MA{Month-Ahead}}}                                                          \\
    Month-Ahead price             & $y_t$                             & 10.027             & 0.056            \\
    Lagged Month-Ahead price      & $y_{t-1}$                         & 7.190              & 0.194            \\
    Lagged Two-Month-Ahead price  & $y_{t-1, \text{MP2}}$             & 6.940              & 0.178            \\
    First day of month dummy      & First day of month                & 0.115              & 0.247            \\
    Days in month                 & $D_t^{M}$                         & 0.999              & 0.280            \\
    Day since last close          & $D_t^{LTD}$                       & 1.281              & 0.146            \\
    Seasonal                      & $\Bar{y}_{t-1Y}^M$                & 4.812              & 0.384            \\
    Lagged EUA price              & $\text{EUA}_{t-1}$                & 14.787             & 0.182            \\
    Lagged oil price              & $\text{Oil}_{t-1}$                & 12.370             & 0.311            \\
    Lagged temperature            & $T_{t-1}$                         & 8.322              & 0.509            \\
    Lagged standardized residual  & $z_{t-1}$                         & 1.008              & 0.663            \\
    \bottomrule
  \end{tabular}
\end{table}
The maximum likelihood estimates themselves are asymptotically normal under some regularity conditions, which often hold in practice. That is, we assume asymptotic normality and conduct z-tests to assess the parameter significance. We approximate the variance-covariance matrix by inverting the Hessian, which is calculated by numerical differentiation using the symmetric difference quotient. The obtained Hessian will be equivalent to the Fisher information matrix, as all models were estimated by minimizing the negative log-likelihood function. Therefore, inverting it yields an estimate for the variance-covariance matrix. We approximated the variance-covariance matrix in every iteration step of the forecasting study and averaged afterward. This makes the estimated Hessian more robust, and the standard errors are not as low as they would be when only evaluating the last step of the expanding window study.

\Cref{tab:param_res_da} summarizes the parameter estimates of the proposed Day-Ahead model along with their significance estimates. The smoothing parameter $\lambda$ is significantly different from one, which indicates that the conditional mean process is not purely a random walk. The seasonality parameter is insignificant, although it reduces the out-of-sample \ac{CRPS} (see \Cref{appendix_da}).
The coal and \ac{EUA} parameters are significant, while the power parameter only reaches weak significance. The ARMA(1,1) process of the \ac{ETS} model is highly significant, indicating that an exponential smoothing model alone does not suffice in describing the process. All parameters of the \ac{TGARCH} process are significant. The high test statistics of the \ac{AR} and \ac{MA} terms suggest that a higher-order \ac{TGARCH} model might be even better.
The skewness parameter $\nu$ is not significantly different from one. A value of one corresponds to asymmetric distribution. The estimated degrees of freedom $\tau$ are significantly higher than four. We can conclude that the fourth moment exists with high certainty since the degrees of freedom function as a tail-index estimate for the T-distribution. This property is important because it renders inference on the \ac{TGARCH} parameters, for which the fourth moment is needed, plausible.

The positive signs of coal and oil prices confirm previous findings in related research, as discussed in \Cref{introduction}. The effect of the \ac{EUA} was estimated to be significantly positive as well, which indicates that the Day-Ahead product is used to substitute coal for power production.

\begin{table}[h!]
  \centering
  \caption{\label{tab:param_res_da}Parameter estimates and test statistics of corresponding z-tests of the proposed Day-Ahead model along with significance levels indicated by colors}
  \begin{tabular}{cccrcr}
    \toprule
    Process  & Parameter        & Description   & Estimate & $\mathcal{H}_0$      & Z-Statistic                               \\
    \midrule
    $\mu$    & $\lambda$        & ETS Smoothing & 1.152    & $\lambda = 1$        & \cellcolor{ssss} 6.65                     \\
    $\mu$    & $\psi_0$         & Seasonal      & -0.015   & $\psi_0=0$           & -1.29                                     \\
    $\mu$    & $\varphi_1$      & AR1           & -0.685   & $\varphi_1 = 0$      & \cellcolor{ssss}  -36.17                  \\
    $\mu$    & $\theta_1$       & MA1           & 0.413    & $\theta_1 = 0$       & \cellcolor{ssss} 16.84                    \\
    $\mu$    & $\zeta_0$        & Coal          & 0.019    & $\zeta_0 = 0$        & \cellcolor{ssss} 3.70                     \\
    $\mu$    & $\zeta_1$        & EUA           & 0.078    & $\zeta_1 = 0$        & \cellcolor{ssss} 3.92                     \\
    $\mu$    & $\zeta_2$        & Power         & 0.001    & $\zeta_2 = 0$        & \cellcolor{s} 1.72                        \\
    $\sigma$ & $\omega$         & Constant      & 0.025    & $\omega = 0$         & \cellcolor{ssss} 15.52                    \\
    $\sigma$ & $\alpha$         & AR1           & 0.291    & $\alpha = 0$         & \cellcolor{ssss} 38.98                    \\
    $\sigma$ & $\beta$          & MA1           & 0.726    & $\beta = 0$          & \cellcolor{ssss} 128.60                   \\
    $\sigma$ & $\alpha + \beta$ & Persistence   & 1.027    & $\alpha + \beta = 1$ & 0.85                                      \\
    $\sigma$ & $\gamma$         & Leverage      & -0.101   & $\gamma = 0$         & \cellcolor{ssss} -9.38                    \\
    $\sigma$ & $\delta$         & Monday        & 1.342    & $\delta = 1$         & \cellcolor{ssss}  5.05                    \\
    $\nu$    & $\nu$            & Nu            & 1.039    & $\nu = 1$            & 1.05                                      \\
    $\tau$   & $\tau$           & Tau           & 6.425    & $\tau > 4$           & \cellcolor{sss}2.91                       \\
    \bottomrule
    \multicolumn{4}{l}{\footnotesize \colorbox{s}{$p\leq0.10$}, \colorbox{sss}{$p\leq0.01$}, \colorbox{ssss}{$p \leq 0.001$}} \\
  \end{tabular}
\end{table}
The $\alpha$ and $\beta$ (the persistence parameters in the \ac{TGARCH} process) sum up to $1.027$. Conducting a z-test with $\mathcal{H}_0: \alpha + \beta = 1$ yields test statistic of $0.85$. That is, $\alpha + \beta$ is not significantly different from $1$. The latter is a characteristic property of \ac{IGARCH} processes. Shocks have an infinite persistence in the \ac{IGARCH} process. In consequence, long term forecasts of the conditional variance will not converge towards the unconditional variance, which is undefined in the \ac{IGARCH} model. Therefore, forecasts will get less reliable as the forecasting horizon increases \autocite{tsay2010analysis}. \ac{IGARCH} behavior can be caused by unaccounted structural change in the conditional variance \autocite{morana2002igarch}. However, if this applies without modifications to the proposed \ac{TGARCH} model is questionable since it deviates from the plain \ac{GARCH} model in various aspects. However, we did estimate the proposed model with $\alpha + \beta$ fixed to $1$. This did not improve the forecasting performance (see \Cref{appendix_da}). Therefore, further examining this issue remains a task for future scientific work.

\begin{table}[h!]
  \centering
  \caption{\label{tab:param_res_ma}Parameter estimates and test statistics of corresponding z-tests of the proposed Month-Ahead model, along with significance levels indicated by colors}
  \begin{tabular}{cccrcr}
    \toprule
    Process  & Parameter        & Description & Estimate & H0                   & Z-Statistic                                 \\
    \midrule
    $\mu$    & $\eta$           & Risk Premia & -0.353   & $\eta = 0$           & \cellcolor{ssss} -3.36                      \\
    $\mu$    & $\psi_0$         & Seasonal    & -0.005   & $\psi_0=0$           & \cellcolor{ssss}   -19.29                   \\
    $\mu$    & $\varphi_0$      & Constant    & 0.112    & $\varphi_0 = 0$      & \cellcolor{ssss} 24.78                      \\
    $\mu$    & $\varphi_1$      & AR1         & 0.997    & $\varphi_1 = 1$      & \cellcolor{ssss} -11.38                     \\
    $\mu$    & $\zeta_0$        & EUA         & -0.005   & $\zeta_0 = 0$        & \cellcolor{ssss}      -6.45                 \\
    $\mu$    & $\zeta_1$        & Oil         & 0.001    & $\zeta_1 = 0$        & \cellcolor{ssss} 16.19                      \\
    $\mu$    & $\zeta_2$        & Temperature & 0.001    & $\zeta_2 = 0$        & \cellcolor{sss} 3.18                        \\
    $\sigma$ & $\omega$         & Constant    & 0.012    & $\omega = 0$         & \cellcolor{ssss}    13.23                   \\
    $\sigma$ & $\alpha$         & AR1         & 0.208    & $\alpha = 0$         & \cellcolor{ssss}   43.27                    \\
    $\sigma$ & $\beta$          & MA1         & 0.834    & $\beta = 0$          & \cellcolor{ssss}      241.51                \\
    $\sigma$ & $\alpha + \beta$ & Persistence & 1.043    & $\alpha + \beta = 1$ & \cellcolor{sss} 2.59                        \\
    $\sigma$ & $\gamma$         & Leverage    & -0.085   & $\gamma = 0$         & \cellcolor{ssss}     -10.47                 \\
    $\sigma$ & $\delta$         & First Day   & 1.403    & $\delta = 1$         & \cellcolor{sss} 2.73                        \\
    $\nu$    & $\nu$            & Nu          & 1.132    & $\nu = 1$            & \cellcolor{ssss} 3.38                       \\
    $\tau$   & $\tau$           & Tau         & 6.858    & $\tau > 4$           & \cellcolor{sss} 3.03                        \\
    \bottomrule
    \multicolumn{4}{l}{\footnotesize \colorbox{s}{$p\leq0.10$}, \colorbox{sss}{$p\leq0.01$}, \colorbox{ssss}{$p \leq 0.001$}} \\
  \end{tabular}
\end{table}
\Cref{tab:param_res_ma} presents the parameter estimates and test statistics of the proposed Month-Ahead model.
The risk parameter is negative as expected, and it is highly significant.
The \ac{EUA} parameter is significantly negative, which contrasts the positive estimate that was observed in the Day-Ahead model. This potentially indicates that the Month-Ahead product is not used for substitution but is rather influenced on its own. While this is a plausible explanation, one has to keep in mind that \textit{Day-Ahead} \ac{EUA} prices were used because other time series of \ac{EUA} prices were not available. Analyzing these effects in more detail remains a topic for further research. However, while the sign of oil is plausible and matches the one from the Day-Ahead model, the sign of the temperature is not plausible. Lower temperatures should lead to a higher natural gas demand, and higher demand usually translated to higher prices if it is not fully anticipated upfront. The sign of the seasonal parameter is also unexpectedly negative. All parameters in the \ac{TGARCH} process are significant. The $\delta$ estimates are significant around 1.4, indicating that the conditional variance is significantly increased on the first trading day of every month. The estimated degrees of freedom $\tau$ are larger than four as for the Day-Ahead model.

\subsection{Predictive Performance}

\Cref{tab:da_ma_performance} shows the forecasting performance of the Day-Ahead and Month-Ahead models. The proposed models outperform their closest benchmarks by $\approx 13$\% and $\approx 9$\% in terms of \ac{CRPS}, respectively. The assumption of T-distributed errors is beneficial for all Day-Ahead models except the Random Forest.  Interestingly, the normality assumption yields better performance for the benchmark models. The results further support the hypothesis that Oil Prices contain predictive power as the \ac{VAR} model consistently outperforms the \ac{ARIMA} counterpart. However, the results of the Random Forest are not in line with the results of \citeauthor{herrera2019long} where the Random Forest substantially outperformed all other considered models.
\begin{table}[h!]
  \centering
  \caption{\label{tab:da_ma_performance}Predictive performance of the proposed models and benchmarks. Lower values indicate better performance.}
  \begin{tabular}{llccc@{}}
    \toprule
                           & \textbf{Reference}            & \textbf{MAE} & \textbf{RMSE} & \textbf{CRPS} \\ \midrule
    \multicolumn{4}{@{}l}{\textbf{\DA{Day-Ahead}}}                                                        \\
    Proposed Model         &                               & 0.3863       & 1.0843        & 0.2834        \\
    VAR (T)                &                               & 0.4144       & 1.1598        & 0.3252        \\
    ARIMA(2,1,2) (T)       &                               & 0.4215       & 1.2127        & 0.3324        \\
    VAR (Normal)           & \cite{geng2017relationship}   & 0.4144       & 1.1598        & 0.3386        \\
    ARIMA(2,1,2) (Normal)  & \cite{siddiqui2019predicting} & 0.4215       & 1.2127        & 0.3466        \\
    Random Forest (Normal) & \cite{herrera2019long}        & 0.5725       & 1.2548        & 0.4361        \\
    Random Forest (T)      &                               & 0.5725       & 1.2548        & 0.4523        \\
    \midrule
    \multicolumn{4}{@{}l}{\textbf{\MA{Month-Ahead}}}                                                      \\
    Proposed Model         &                               & 0.3010       & 0.3995        & 0.2126        \\
    VAR (Normal)           & \cite{geng2017relationship}   & 0.3184       & 0.4619        & 0.2336        \\
    ARIMA(2,1,2) (Normal)  & \cite{siddiqui2019predicting} & 0.3201       & 0.4635        & 0.2349        \\
    VAR (T)                &                               & 0.3184       & 0.4619        & 0.2356        \\
    ARIMA(2,1,2) (T)       &                               & 0.3201       & 0.4635        & 0.2371        \\
    Random Forest (Normal) & \cite{herrera2019long}        & 0.4681       & 0.6167        & 0.3502        \\
    Random Forest (T)      &                               & 0.4681       & 0.6167        & 0.3603        \\
    \bottomrule
  \end{tabular}
\end{table}

\begin{figure}[h!]
  \centering
  \fbox{\includegraphics[width=0.975\columnwidth]{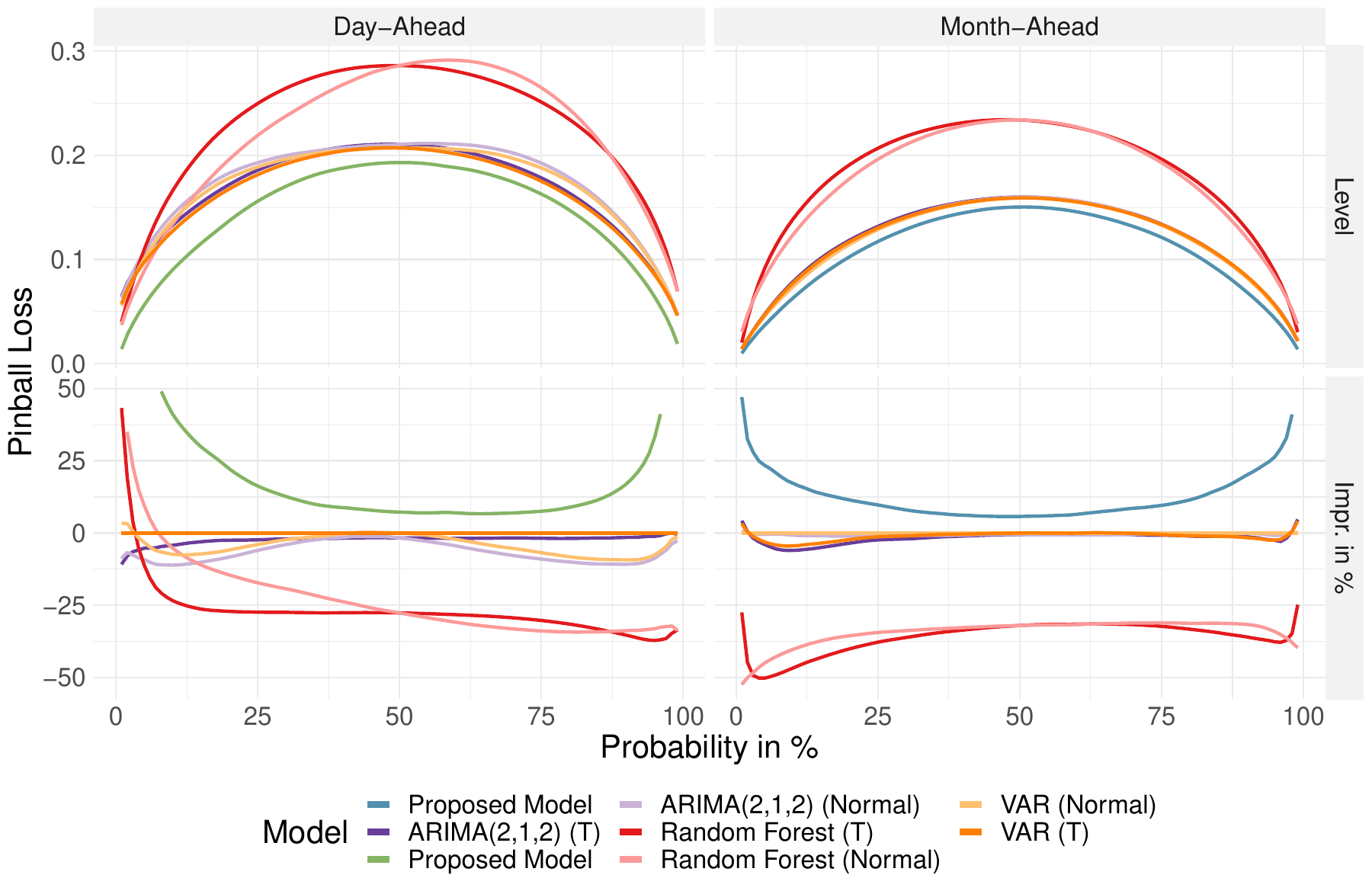}}
  \caption{\label{fig:pinball_loss}Pinball loss levels at all percentiles for the Day-Ahead (top-left) and Month-Ahead (top-right) models as well as the relative improvement compared to the best performing benchmark model - the VAR (T) model for Day-Ahead (bottom-left) and VAR (Normal) for Month-Ahead (bottom-right)}
\end{figure}

However, the results in \Cref{tab:da_ma_performance} do not reveal how these improvements are distributed over the predictive distribution. \Cref{fig:pinball_loss} closes this gap. It shows the Pinball Loss of both proposed models and their benchmarks. The figure presents the level of the Pinball Loss (top), as well as the improvement in \% relative to the best performing benchmark (bottom). Both proposed models outperform their benchmarks over the whole probability grid. Moreover, substantial improvements are realized in the distribution's tails, which are particularly relevant for risk management.

\begin{figure}[h!]
  \centering
  \fbox{\includegraphics[width=0.975\columnwidth]{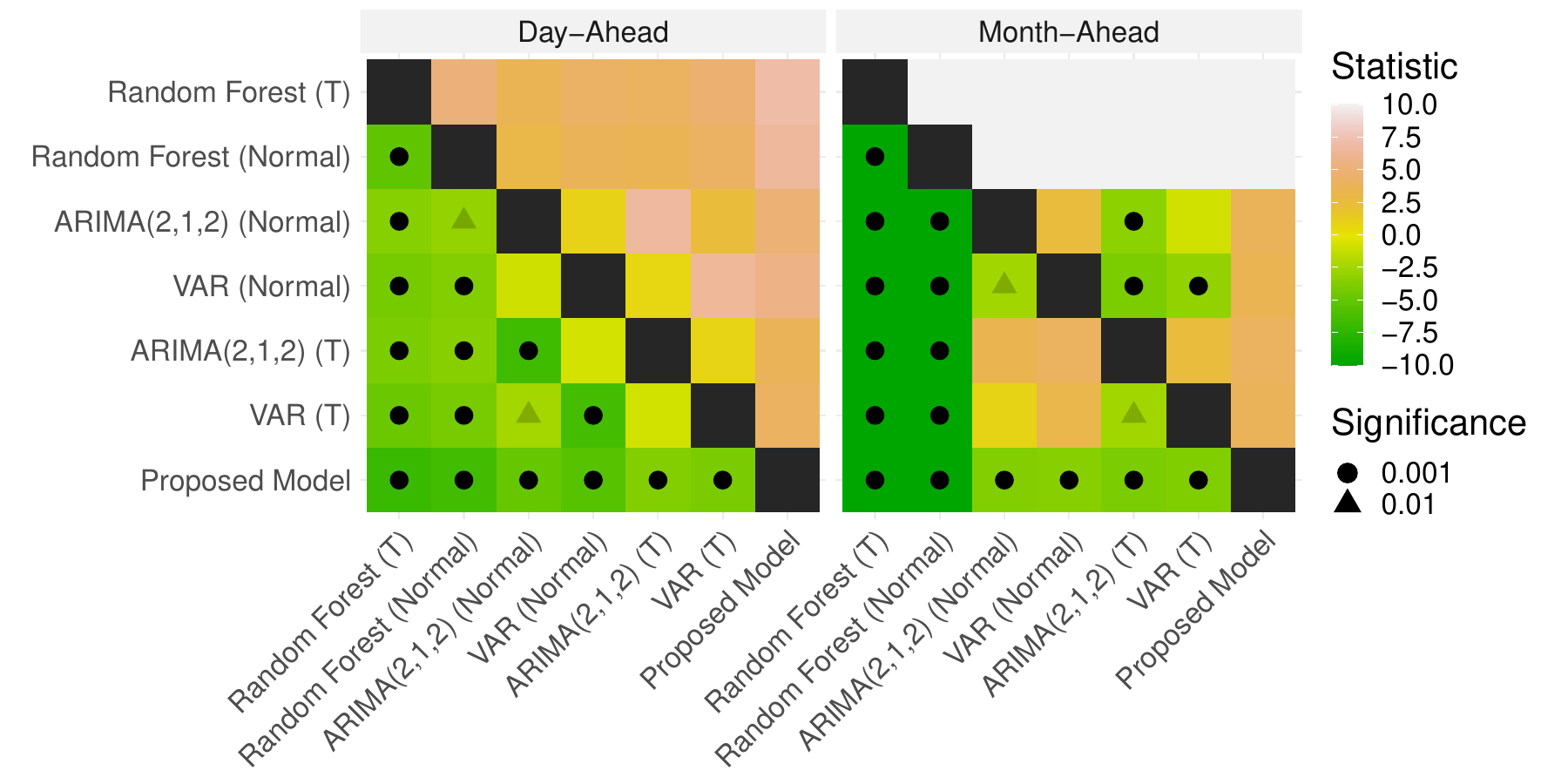}}
  \caption{\label{fig:dm_plot_benches_crps}Results of DM-Tests testing whether the y-axis model performs better than the x-axis model. The colors correspond to levels of the test statistics. Symbols denote selected significance levels.}
\end{figure}

We conducted the DM-Test to check whether the observed improvements in \ac{CRPS} are significant. \Cref{fig:dm_plot_benches_crps} contains the  results. The colors indicate the value of the test statistic, while the symbols note significance. Lower values (dark green) indicate that the model on the y-axis outperforms the model on the x-axis. All models outperform the Random Forest, while the \ac{VAR} outperforms most other models for both products. The proposed models significantly outperform all benchmarks, which makes them preferable.

\begin{figure}[h!]
  \centering
  \fbox{\includegraphics[width=0.975\columnwidth]{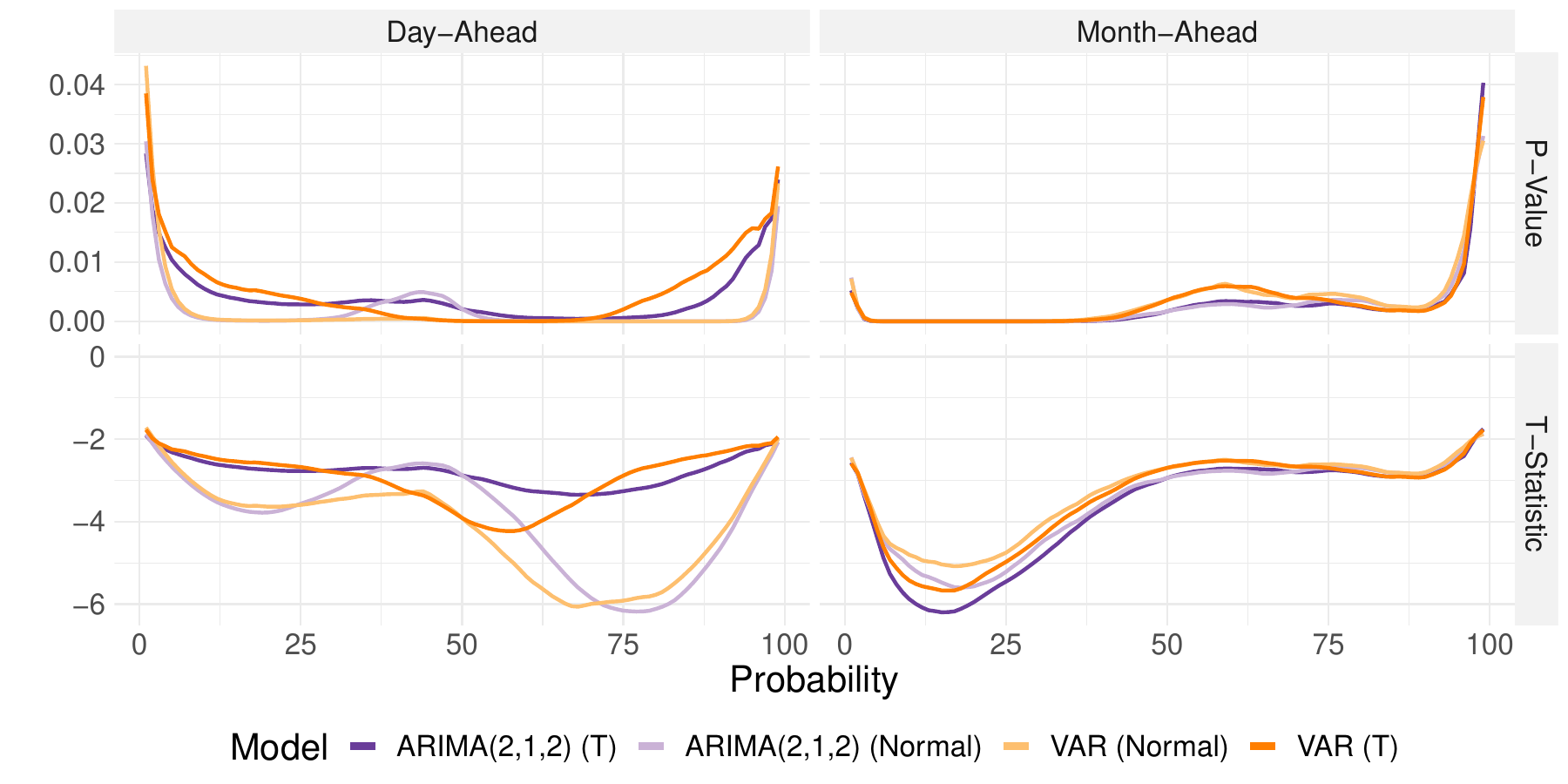}}
  \caption{\label{fig:pb_dm_plot}Test statistics (top) and p-values (bottom) of DM tests testing whether the predictions of the proposed models are better than those from the ARIMA and VAR models. Lower values indicate better performance of the proposed models.}
\end{figure}

We also tested if the observed differences in Pinball Loss (see \Cref{fig:pinball_loss}) are significant. Therefore we used the DM-Test with respect to the Pinball Loss of each percentile. That is, we can present the results concisely in \Cref{fig:pb_dm_plot}. It shows the P-Values (top) and test-statistics (bottom). Lower values indicate that the proposed model outperforms the benchmark. Furthermore, all P-Values are below 5\% for all benchmarks. This means that the proposed models significantly outperform the benchmarks at all percentiles.

\begin{figure}[h!]
  \centering
  \fbox{\includegraphics[width=0.975\columnwidth]{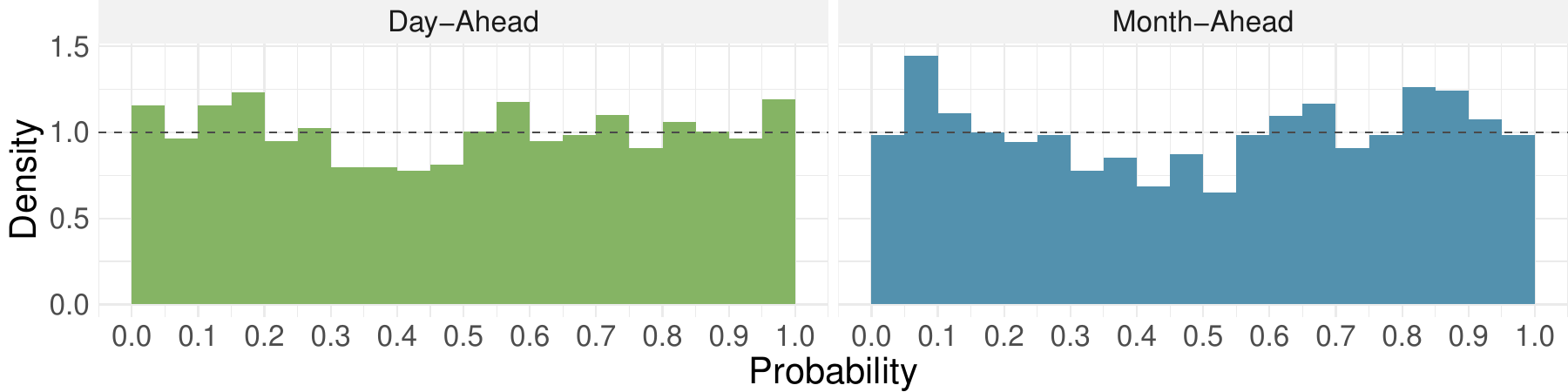}}
  \caption{\label{fig:pit}PIT histogram of the probabilistic out-of-sample predictions of both proposed models}
\end{figure}

Lastly, we constructed a probability integral transform (PIT) histogram for both proposed models to evaluate the calibration accuracy, see \Cref{fig:pit}.
Each bin represents a probability segment of the predictive distribution with a predicted probability of 5\%. A perfectly calibrated model would produce a uniformly shaped histogram. That is, the figure indicates that both models are calibrated reasonably well while the Month-Ahead model has a slight tendency of underdispersion \autocite{raftery2005using}.

\begin{figure}[h!]
  \centering
  \fbox{\includegraphics[width=0.975\columnwidth]{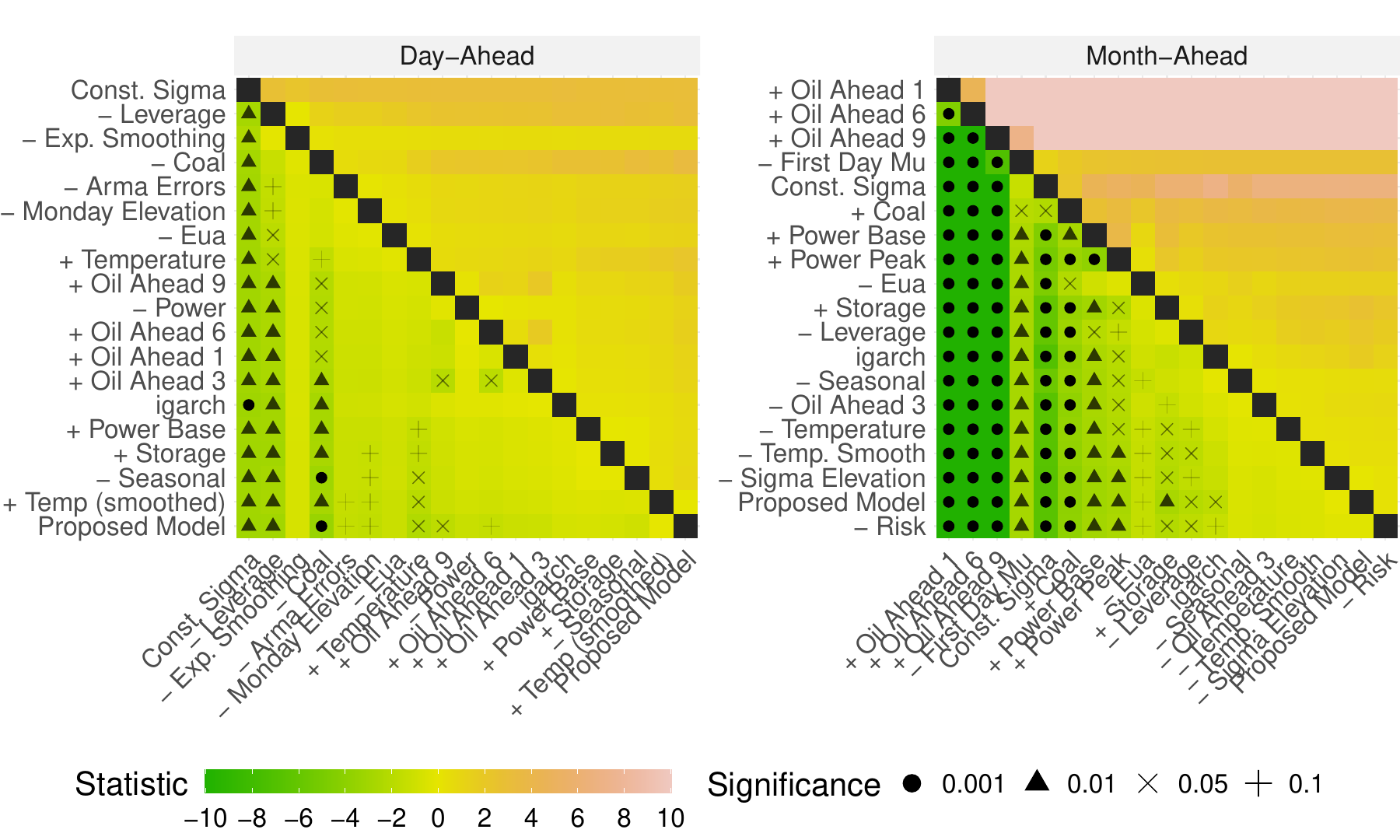}}
  \caption{\label{fig:dm_plot}Results of DM-Tests with respect to CRPS, testing whether the y-axis model performs better than the x-axis model. Colors denote the level of the test statistics. Symbols indicate selected significance levels.}
\end{figure}

To keep this paper concise, we decided to present the proposed model only instead of presenting each model considered during the model building process.
We only discussed the proposed models in detail to keep this paper concise. However, we considered several variations of the two proposed models, either leaving out (-) or adding (+) model components. This is helpful to assess the relative importance of each component. We used the DM-Test with respect to the CRPS to compare the predictive performance of these models. \Cref{fig:dm_plot} shows the results. The figure can be interpreted as follows. The main body shows the test-statistics of DM-Tests testing whether the y-axis model outperforms the x-axis model. Therefore, inspecting the last column of the Day-Ahead facet answers whether any model (denoted on the y-axis) outperforms the proposed model (the last entry on the x-axis). The latter is not the case which means that neither adding nor removing model components yields a significantly better predictive performance. This also holds true for the proposed Month-Ahead model.

Some findings are worth highlighting here. First, the Models assuming homoscedasticity are among the worst for both products. The Day-Ahead Model profits in particular from including \ac{ARMA} errors, coal, and the leverage effect in the sigma process. The Month-Ahead model profits from considering the rollover (First Day Mu) as well as including \ac{EUA} and the leverage effect. Removing components not always produces a significant difference. For example, removing EUA prices from the proposed Day-Ahead model does not yield a significantly worse performance with respect to the DM-Test. However, the EUA coefficient itself is significant, as shown in \Cref{tab:param_res_da} and including EUA prices reduces the out-of-sample CRPS. Therefore, we keep EUA prices included in the proposed model.

\ac{CRPS}, \ac{RMSE} and \ac{MAE} values of all models are summarized in \Cref{tab:da_delta,tab:ma_delta} in \Cref{deltamodels}. Results of DM-Tests testing with respect to the absolute and squarred error are also included in \Cref{deltamodels} (see \Cref{fig:dm_plot_da,fig:dm_plot_ma}).

\section{Conclusion}\label{conclusion}

In this paper, we examined the problem of forecasting European natural gas prices. We developed state-space time series models for probabilistic price forecasting of the Day-Ahead and Month-Ahead products.
Both time series are heavy-tailed and feature conditional heteroscedasticity with leverage effects. In addition, the Month-Ahead time series contains risk premia effects.
The model development process was guided by an extensive exploratory data analysis (see \Cref{data}) and results from related research.
The econometric models were estimated using maximum likelihood under skewed T-distribution. Out-of-sample forecasts obtained from an expanding window forecasting study were used to evaluate the forecasting performance using various probabilistic performance measures, including the \ac{CRPS} and the Pinball loss.

Particularly, the Day-Ahead time-series features pronounced heavy tails and benefits substantially from assuming a t-distribution.
We modeled the conditional mean using a state-space model with ARMA errors, a seasonal component, and external regressors.
We used a stepwise forward selection approach to determine the influence of external regressors. The results show a significant positive impact on the coal price. The latter confirms previous findings \autocite{papiez2011analysis}.
Additionally, we find a positive influence of \ac{EUA} and power peak prices on natural gas prices. This indicates a strong link between natural gas and electricity markets in Europe. Finally, the conditional standard deviation was modeled using a \ac{TGARCH} process, extended by a level elevation on Mondays. As a result, the estimated volatility is increased by around 34\% on Mondays.
The proposed Day-Ahead model improves the \ac{CRPS} by $12.9\%$ compared to a \ac{VAR} model with t-distributed errors, which is the best benchmark.

A major consideration for the Month-Ahead time series was the presence of risk premia and the rollover problem. The latter implies a sudden change in the delivery period on the first trading day of each month. This leads to pronounced price effects and, as a consequence, increased conditional volatility. The proposed model covers these effects by using the last observed price of the Two-Month-Ahead product as a predictor for the first trading day and by elevating the conditional standard deviation on the first trading day. The latter increase is estimated to be 40\% in the proposed Month-Ahead model. The observed positive changes at the beginning of the month are possibly due to risk premia. These risk premia should decrease over time because the uncertainty decreases as time gets closer to delivery. We find this highly significant pattern in the data estimating a total price reduction of $0.35\EUR$ per month.
Including a seasonal component as well as \ac{EUA} prices, oil prices, and the temperature reduced the \ac{CRPS}. However, only the effect of \ac{EUA} prices is significant. The impact is negative, which might indicate that the Month-Ahead product is not used for substituting coal, in contrast to the Day-Ahead model, which is positively influenced by \ac{EUA} prices. The leverage component in the \ac{GARCH} process is significant and negative in the proposed Month-Ahead model.
Using the proposed Month-Ahead model compared to the \ac{VAR} alternative with Gaussian errors leads to a $9$\% reduction in out of sample \ac{CRPS}.

Some additional questions arise in light of the results. Further analyzing the influence of \ac{EUA} prices seems interesting for energy companies and governments. The main question to be answered here is why the impact of \ac{EUA} prices on natural gas prices is positive for the Day-Ahead product while it is negative for the Month-Ahead product.
Further investigation of the role of risk premia in natural gas markets is another attractive topic since the literature on natural gas price forecasting has not covered it yet.

\printbibliography[title = References]

\cleardoublepage
\appendix

\setcounter{table}{0}
\setcounter{figure}{0}
\renewcommand{\thetable}{A\arabic{table}}
\renewcommand{\thefigure}{A\arabic{figure}}

\section{Results of Reduced Models}\label{deltamodels}

\subsection{Day-Ahead}\label{appendix_da}

\begin{table}[!h]

  \caption{\label{tab:da_delta}Predictive Performance of the proposed Day-Ahead model and variations thereof sorted by the relative difference in CRPS (last column)}
  \centering
  \begin{tabular}[t]{llrrrr}
    \toprule
    Num. & Changed Component  & MAE    & RMSE   & CRPS   & CRPS $\Delta$ (\%) \\
    \midrule
    19   & Proposed Model     & 0.3863 & 1.0843 & 0.2834 & 0.0000             \\
    18   & + Temp (smoothed)  & 0.3866 & 1.0844 & 0.2834 & 0.0015             \\
    17   & - Seasonal         & 0.3866 & 1.0846 & 0.2836 & 0.0495             \\
    16   & + Storage          & 0.3862 & 1.0840 & 0.2836 & 0.0520             \\
    15   & + Power Base       & 0.3861 & 1.0837 & 0.2836 & 0.0528             \\
    14   & IGARCH             & 0.3864 & 1.0843 & 0.2837 & 0.0871             \\
    13   & + Oil Ahead 3      & 0.3869 & 1.0843 & 0.2837 & 0.0951             \\
    12   & + Oil Ahead 1      & 0.3869 & 1.0847 & 0.2837 & 0.1077             \\
    11   & + Oil Ahead 6      & 0.3870 & 1.0847 & 0.2838 & 0.1262             \\
    10   & - Power            & 0.3866 & 1.0835 & 0.2838 & 0.1277             \\
    09   & + Oil Ahead 9      & 0.3870 & 1.0842 & 0.2839 & 0.1549             \\
    08   & + Temperature      & 0.3872 & 1.0842 & 0.2841 & 0.2332             \\
    07   & - Eua              & 0.3865 & 1.0853 & 0.2842 & 0.2857             \\
    06   & - Monday Elevation & 0.3862 & 1.0843 & 0.2844 & 0.3417             \\
    05   & - Arma Errors      & 0.3868 & 1.0868 & 0.2845 & 0.3756             \\
    04   & - Coal             & 0.3889 & 1.0849 & 0.2850 & 0.5677             \\
    03   & - Exp. Smoothing   & 0.3878 & 1.1010 & 0.2860 & 0.9132             \\
    02   & - Leverage         & 0.3865 & 1.0843 & 0.2863 & 1.0169             \\
    01   & Const. Sigma       & 0.3885 & 1.0826 & 0.3053 & 7.7054             \\
    \bottomrule
  \end{tabular}
\end{table}

\begin{figure}[h!]
  \centering
  \fbox{\includegraphics[width=0.975\columnwidth]{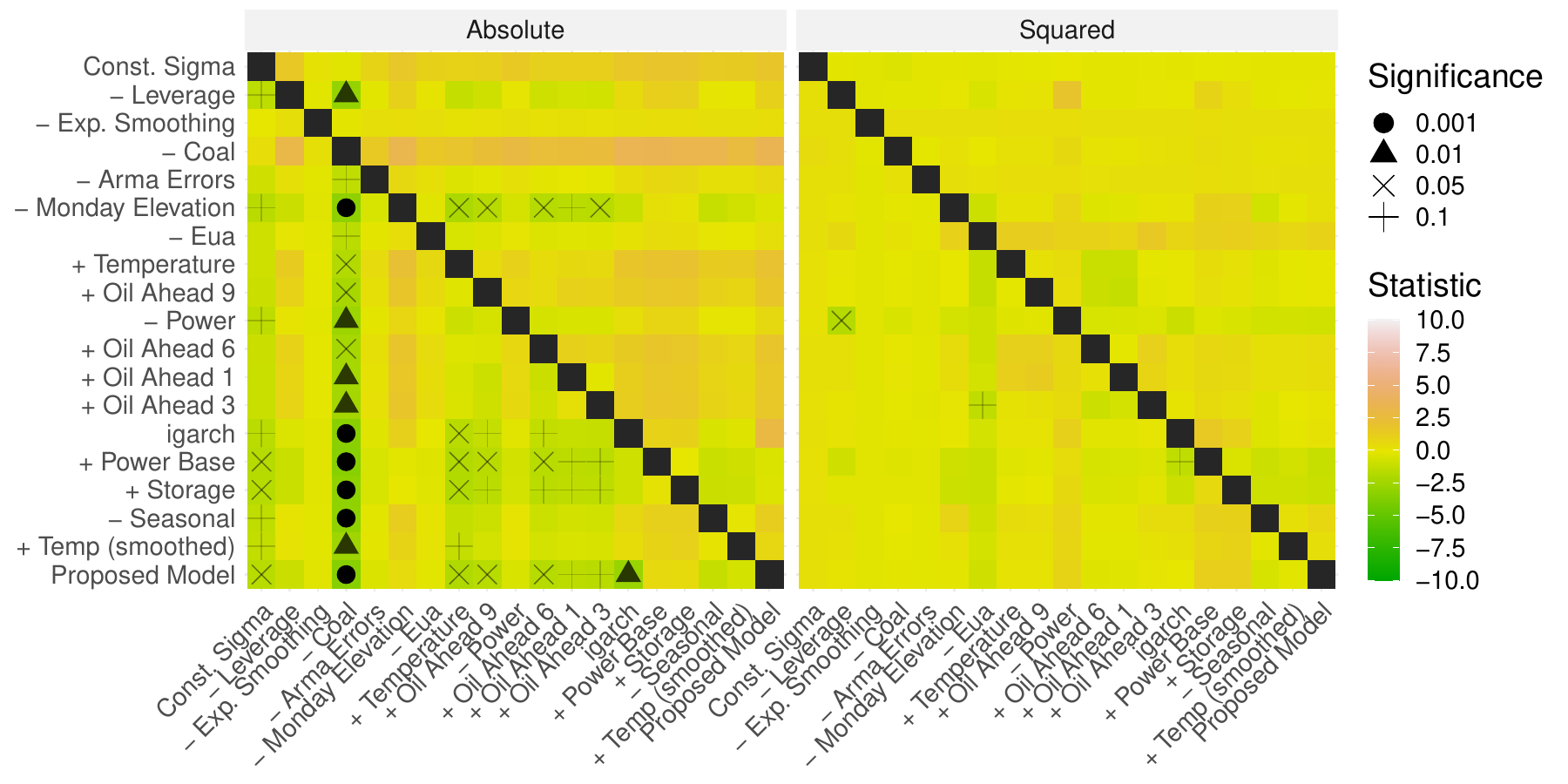}}
  \caption{\label{fig:dm_plot_da}Results of the DM-Test for the proposed Day-Ahead model and variations thereof with respect to absolute and squarred error. The symbols indicate the $.1$, $.05$, $.01$ and $.001$ significance levels.}
\end{figure}

\newpage

\subsection{Month-Ahead}\label{appendix_ma}

\begin{table}[!h]

  \caption{\label{tab:ma_delta}Predictive Performance of the proposed Month-Ahead model and variations thereof sorted by the relative difference in CRPS (last column)}
  \centering
  \begin{tabular}[t]{llrrrr}
    \toprule
    Num. & Changed Component & MAE    & RMSE   & CRPS   & CRPS $\Delta$ (\%) \\
    \midrule
    19   & - Risk            & 0.3011 & 0.3997 & 0.2126 & 0.0238             \\
    18   & Proposed Model    & 0.3010 & 0.3995 & 0.2126 & 0.0000             \\
    17   & - Sigma Elevation & 0.3011 & 0.3996 & 0.2126 & -0.0032            \\
    16   & - Temp. Smooth    & 0.3009 & 0.3997 & 0.2127 & -0.0287            \\
    15   & - Temperature     & 0.3009 & 0.3999 & 0.2127 & -0.0461            \\
    14   & - Oil Ahead 3     & 0.3009 & 0.4000 & 0.2128 & -0.0836            \\
    13   & - Seasonal        & 0.3010 & 0.3996 & 0.2129 & -0.1213            \\
    12   & IGARCH            & 0.3010 & 0.3993 & 0.2129 & -0.1452            \\
    11   & - Leverage        & 0.3011 & 0.3995 & 0.2132 & -0.2744            \\
    10   & + Storage         & 0.3018 & 0.4006 & 0.2133 & -0.3412            \\
    09   & - Eua             & 0.3021 & 0.4010 & 0.2137 & -0.4991            \\
    08   & + Power Peak      & 0.3029 & 0.4011 & 0.2140 & -0.6541            \\
    07   & + Power Base      & 0.3036 & 0.4018 & 0.2146 & -0.9087            \\
    06   & + Coal            & 0.3073 & 0.4029 & 0.2167 & -1.9002            \\
    05   & Const. Sigma      & 0.3017 & 0.4009 & 0.2205 & -3.7067            \\
    04   & - First Day Mu    & 0.3170 & 0.4617 & 0.2276 & -7.0538            \\
    03   & + Oil Ahead 9     & 0.3861 & 0.5133 & 0.2839 & -33.5322           \\
    02   & + Oil Ahead 6     & 0.5194 & 0.6788 & 0.3984 & -87.3583           \\
    01   & + Oil Ahead 1     & 0.6173 & 0.7443 & 0.4699 & -120.9902          \\
    \bottomrule
  \end{tabular}
\end{table}

\begin{figure}[h!]
  \centering
  \fbox{\includegraphics[width=0.975\columnwidth]{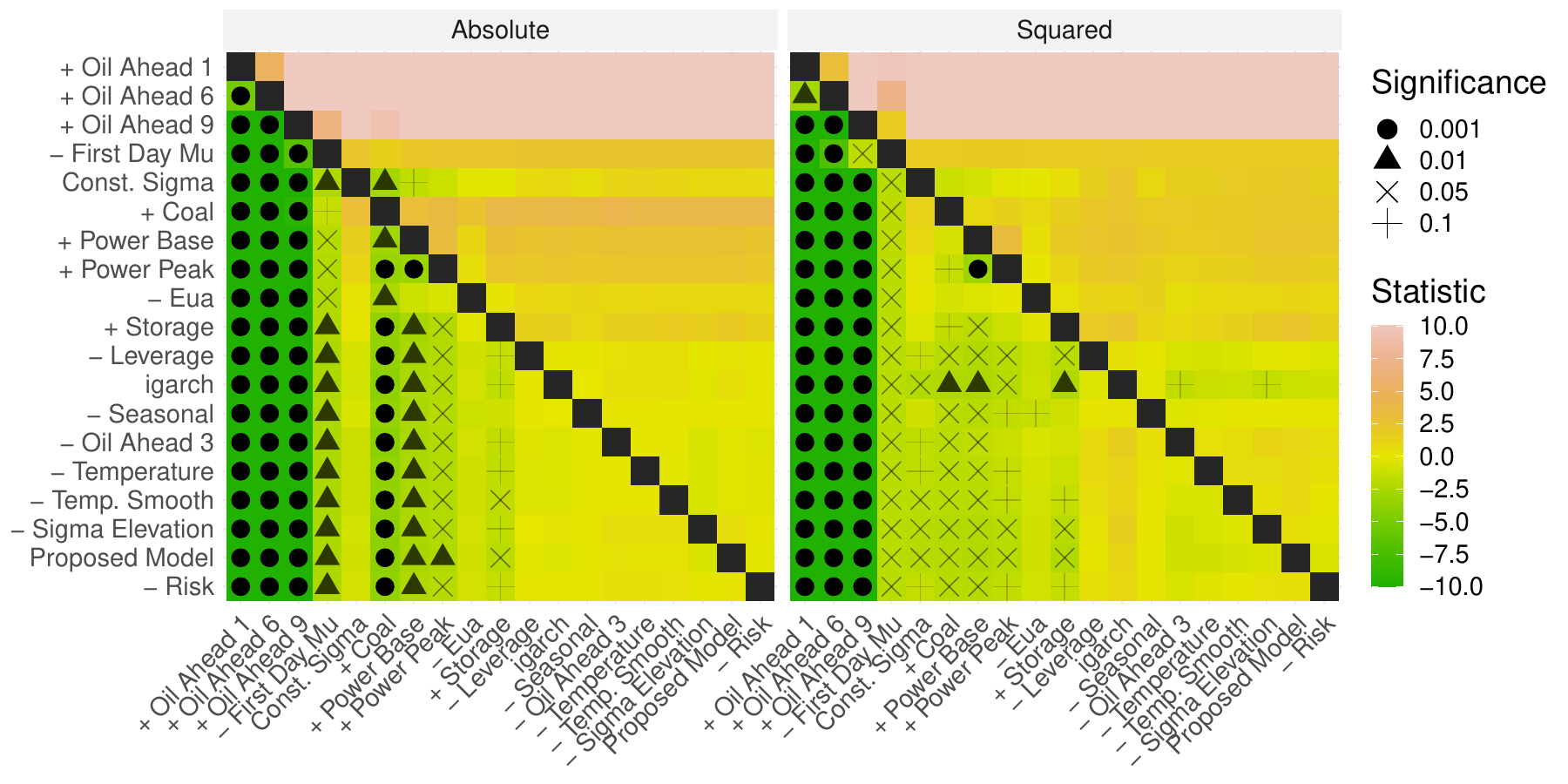}}
  \caption{\label{fig:dm_plot_ma} Results of the DM-Test for the proposed model and variations thereof with respect to absolute and squarred error. The symbols indicate the $.1$, $.05$, $.01$ and $.001$ significance levels.}
\end{figure}

\newpage

\end{document}